\documentclass[twocolumn]{aastex63}

\usepackage{graphicx}
\usepackage[T1]{fontenc}
\usepackage{aecompl}
\usepackage[]{amsmath}

\usepackage{color}
\usepackage{float}
\usepackage{multirow}

\usepackage{xspace}
\usepackage{soul}
\usepackage{placeins}
\usepackage{atbegshi}
\newcommand{\handlethispage}{}
\newcommand{\discardpagesfromhere}{\let\handlethispage\AtBeginShipoutDiscard}
\newcommand{\keeppagesfromhere}{\let\handlethispage\relax}
\usepackage{lineno}

\graphicspath{{figures/}}

\def\lsim{ \lower .75ex \hbox{$\sim$} \llap{\raise .27ex \hbox{$<$}} }

\submitjournal{ApJ}

\shorttitle{Dark matter content of Draco, Sextans and Ursa Minor}
\shortauthors{Yang et al.}

\graphicspath{{./}{figures/}}

\begin{document}

\title{The dark matter content of Milky Way dwarf spheroidal galaxies: Draco, Sextans and Ursa Minor}


\author{Hao Yang}
\thanks{haoyang22@sjtu.edu.cn}
\affiliation{Department of Astronomy, School of Physics and Astronomy, and Shanghai Key Laboratory for Particle Physics and Cosmology, Shanghai Jiao Tong University, Shanghai 200240, People's Republic of China}
\affiliation{Tsung-Dao Lee Institute, Shanghai Jiao Tong University, Shanghai, 201210, China}
\affiliation{State Key Laboratory of Dark Matter Physics, School of Physics and Astronomy,Shanghai Jiao Tong University, Shanghai 200240, China}
\author[0000-0002-5762-7571]{Wenting Wang}
\thanks{Corresponding author: wenting.wang@sjtu.edu.cn}
\affiliation{Department of Astronomy, School of Physics and Astronomy, and Shanghai Key Laboratory for Particle Physics and Cosmology, Shanghai Jiao Tong University, Shanghai 200240, People's Republic of China}
\affiliation{State Key Laboratory of Dark Matter Physics, School of Physics and Astronomy,Shanghai Jiao Tong University, Shanghai 200240, China}
\author{Ling Zhu}
\affiliation{Shanghai Astronomical Observatory, Chinese Academy of Sciences, 80 Nandan Road, Shanghai 200030, China}
\author{Ting S. Li}
\affiliation{Department of Astronomy and Astrophysics, University of Toronto, 50 St. George Street, Toronto ON, M5S 3H4, Canada}
\author[0000-0003-2644-135X]{Sergey E. Koposov}
\affiliation{Institute for Astronomy, University of Edinburgh, Royal Observatory, Blackford Hill, Edinburgh EH9 3HJ, UK}
\affiliation{Institute of Astronomy, University of Cambridge, Madingley Road, Cambridge CB3 0HA, UK}
\author{Jiaxin Han}
\affiliation{Department of Astronomy, School of Physics and Astronomy, and Shanghai Key Laboratory for Particle Physics and Cosmology, Shanghai Jiao Tong University, Shanghai 200240, People's Republic of China}
\affiliation{State Key Laboratory of Dark Matter Physics, School of Physics and Astronomy,Shanghai Jiao Tong University, Shanghai 200240, China}
\author[0000-0002-6469-8263]{Songting Li}
\affiliation{Department of Astronomy, School of Physics and Astronomy, and Shanghai Key Laboratory for Particle Physics and Cosmology, Shanghai Jiao Tong University, Shanghai 200240, People's Republic of China}
\affiliation{State Key Laboratory of Dark Matter Physics, School of Physics and Astronomy,Shanghai Jiao Tong University, Shanghai 200240, China}
\author{Rui Shi}
\affiliation{Department of Astronomy, School of Physics and Astronomy, and Shanghai Key Laboratory for Particle Physics and Cosmology, Shanghai Jiao Tong University, Shanghai 200240, People's Republic of China}
\affiliation{State Key Laboratory of Dark Matter Physics, School of Physics and Astronomy,Shanghai Jiao Tong University, Shanghai 200240, China}
\author{Monica Valluri}
\affiliation{Department of Astronomy, University of Michigan, Ann Arbor, MI 48109, USA}
\affiliation{University of Michigan, 500 S. State Street, Ann Arbor, MI 48109, USA}
\author[0000-0001-5805-5766]{Alexander H.~Riley}
\affiliation{Institute for Computational Cosmology, Department of Physics, Durham University, South Road, Durham DH1 3LE, UK}
\author[0000-0002-4928-4003]{Arjun Dey}
\affiliation{NSF NOIRLab, 950 N. Cherry Ave., Tucson, AZ 85719, USA}
\author[0000-0002-6667-7028]{Constance Rockosi}
\affiliation{Department of Astronomy and Astrophysics, UCO/Lick Observatory, University of California, 1156 High Street, Santa Cruz, CA 95064, USA}
\affiliation{Department of Astronomy and Astrophysics, University of California, Santa Cruz, 1156 High Street, Santa Cruz, CA 95065, USA}
\affiliation{University of California Observatories, 1156 High Street, Sana Cruz, CA 95065, USA}
\author{Carles G. Palau}
\affiliation{Department of Astronomy, School of Physics and Astronomy, and Shanghai Key Laboratory for Particle Physics and Cosmology, Shanghai Jiao Tong University, Shanghai 200240, People's Republic of China}
\affiliation{State Key Laboratory of Dark Matter Physics, School of Physics and Astronomy,Shanghai Jiao Tong University, Shanghai 200240, China}
\author{Jessica Nicole Aguilar}
\affiliation{Lawrence Berkeley National Laboratory, 1 Cyclotron Road, Berkeley, CA 94720, USA}
\author[0000-0001-6098-7247]{Steven Ahlen}
\affiliation{Department of Physics, Boston University, 590 Commonwealth Avenue, Boston, MA 02215 USA}
\author{David Brooks}
\affiliation{Department of Physics \& Astronomy, University College London, Gower Street, London, WC1E 6BT, UK}
\author{Todd Claybaugh}
\affiliation{Lawrence Berkeley National Laboratory, 1 Cyclotron Road, Berkeley, CA 94720, USA}
\author[0000-0001-8274-158X]{Andrew Cooper}
\affiliation{Institute of Astronomy and Department of Physics, National Tsing Hua University, 101 Kuang-Fu Rd. Sec. 2, Hsinchu 30013, Taiwan}
\author[0000-0002-1769-1640]{Axel de la Macorra}
\affiliation{Instituto de F\'{\i}sica, Universidad Nacional Aut\'{o}noma de M\'{e}xico,  Circuito de la Investigaci\'{o}n Cient\'{\i}fica, Ciudad Universitaria, Cd. de M\'{e}xico  C.~P.~04510,  M\'{e}xico}
\author{Peter Doel}
\affiliation{Department of Physics \& Astronomy, University College London, Gower Street, London, WC1E 6BT, UK}
\author[0000-0003-4992-7854]{Simone Ferraro}
\affiliation{Lawrence Berkeley National Laboratory, 1 Cyclotron Road, Berkeley, CA 94720, USA}
\affiliation{University of California, Berkeley, 110 Sproul Hall \#5800 Berkeley, CA 94720, USA}
\author[0000-0002-2890-3725]{Jaime E. Forero-Romero}
\affiliation{Departamento de F\'isica, Universidad de los Andes, Cra. 1 No. 18A-10, Edificio Ip, CP 111711, Bogot\'a, Colombia}
\affiliation{Observatorio Astron\'omico, Universidad de los Andes, Cra. 1 No. 18A-10, Edificio H, CP 111711 Bogot\'a, Colombia}
\author{Enrique Gaztañaga}
\affiliation{Institut d'Estudis Espacials de Catalunya (IEEC), c/ Esteve Terradas 1, Edifici RDIT, Campus PMT-UPC, 08860 Castelldefels, Spain}
\affiliation{Institute of Cosmology and Gravitation, University of Portsmouth, Dennis Sciama Building, Portsmouth, PO1 3FX, UK}
\affiliation{Institute of Space Sciences, ICE-CSIC, Campus UAB, Carrer de Can Magrans s/n, 08913 Bellaterra, Barcelona, Spain}
\author[0000-0003-3142-233X]{Satya Gontcho A Gontcho}
\affiliation{Lawrence Berkeley National Laboratory, 1 Cyclotron Road, Berkeley, CA 94720, USA}
\author[0000-0003-4089-6924]{Alma Xochitl Gonzalez Morales}
\affiliation{Departamento de F\'{\i}sica, DCI-Campus Le\'{o}n, Universidad de Guanajuato, Loma del Bosque 103, Le\'{o}n, Guanajuato C.~P.~37150, M\'{e}xico}
\author{Gaston Gutierrez}
\affiliation{Fermi National Accelerator Laboratory, PO Box 500, Batavia, IL 60510, USA}
\author[0000-0001-9822-6793]{Julien Guy}
\affiliation{Lawrence Berkeley National Laboratory, 1 Cyclotron Road, Berkeley, CA 94720, USA}
\author[0000-0002-6550-2023]{Klaus Honscheid}
\affiliation{Center for Cosmology and AstroParticle Physics, The Ohio State University, 191 West Woodruff Avenue, Columbus, OH 43210, USA}
\affiliation{Department of Physics, The Ohio State University, 191 West Woodruff Avenue, Columbus, OH 43210, USA}
\affiliation{The Ohio State University, Columbus, 43210 OH, USA}
\author[0000-0002-6024-466X]{Mustapha Ishak}
\affiliation{Department of Physics, The University of Texas at Dallas, 800 W. Campbell Rd., Richardson, TX 75080, USA}
\author[0000-0003-0201-5241]{Dick Joyce}
\affiliation{NSF NOIRLab, 950 N. Cherry Ave., Tucson, AZ 85719, USA}
\author{Robert Kehoe}
\affiliation{Department of Physics, Southern Methodist University, 3215 Daniel Avenue, Dallas, TX 75275, USA}
\author[0000-0003-3510-7134]{Theodore Kisner}
\affiliation{Lawrence Berkeley National Laboratory, 1 Cyclotron Road, Berkeley, CA 94720, USA}
\author{Namitha Kizhuprakkat}
\affiliation{Institute of Astronomy and Department of Physics, National Tsing Hua University, 101 Kuang-Fu Rd. Sec. 2, Hsinchu 30013, Taiwan}
\affiliation{Center for Informatics and Computation in Astronomy, NTHU, 101 Kuang-Fu Rd. Sec. 2, Hsinchu 30013, Taiwan}
\author[0000-0001-6356-7424]{Anthony Kremin}
\affiliation{Lawrence Berkeley National Laboratory, 1 Cyclotron Road, Berkeley, CA 94720, USA}
\author{Ofer Lahav}
\affiliation{Department of Physics \& Astronomy, University College London, Gower Street, London, WC1E 6BT, UK}
\author[0000-0003-1838-8528]{Martin Landriau}
\affiliation{Lawrence Berkeley National Laboratory, 1 Cyclotron Road, Berkeley, CA 94720, USA}
\author[0000-0001-7178-8868]{Laurent Le Guillou}
\affiliation{Sorbonne Universit\'{e}, CNRS/IN2P3, Laboratoire de Physique Nucl\'{e}aire et de Hautes Energies (LPNHE), FR-75005 Paris, France}
\author[0000-0003-0105-9576]{Gustavo E. Medina}
\affiliation{Department of Astronomy \& Astrophysics, University of Toronto, Toronto, ON M5S 3H4, Canada}
\author[0000-0002-1125-7384]{Aaron Meisner}
\affiliation{NSF NOIRLab, 950 N. Cherry Ave., Tucson, AZ 85719, USA}
\author{Ramon Miquel}
\affiliation{Instituci\'{o} Catalana de Recerca i Estudis Avan\c{c}ats, Passeig de Llu\'{\i}s Companys, 23, 08010 Barcelona, Spain}
\affiliation{Institut de F\'{i}sica d’Altes Energies (IFAE), The Barcelona Institute of Science and Technology, Edifici Cn, Campus UAB, 08193, Bellaterra (Barcelona), Spain}
\author[0000-0003-3188-784X]{Nathalie Palanque-Delabrouille}
\affiliation{IRFU, CEA, Universit\'{e} Paris-Saclay, F-91191 Gif-sur-Yvette, France}
\affiliation{Lawrence Berkeley National Laboratory, 1 Cyclotron Road, Berkeley, CA 94720, USA}
\author[0000-0001-7145-8674]{Francisco Prada}
\affiliation{Instituto de Astrof\'{i}sica de Andaluc\'{i}a (CSIC), Glorieta de la Astronom\'{i}a, s/n, E-18008 Granada, Spain}
\author[0000-0001-6979-0125]{Ignasi Pérez-Ràfols}
\affiliation{Departament de F\'isica, EEBE, Universitat Polit\`ecnica de Catalunya, c/Eduard Maristany 10, 08930 Barcelona, Spain}
\author{Graziano Rossi}
\affiliation{Department of Physics and Astronomy, Sejong University, 209 Neungdong-ro, Gwangjin-gu, Seoul 05006, Republic of Korea}
\author[0000-0002-9646-8198]{Eusebio Sanchez}
\affiliation{CIEMAT, Avenida Complutense 40, E-28040 Madrid, Spain}
\author{David Schlegel}
\affiliation{Lawrence Berkeley National Laboratory, 1 Cyclotron Road, Berkeley, CA 94720, USA}
\author{Michael Schubnell}
\affiliation{Department of Physics, University of Michigan, 450 Church Street, Ann Arbor, MI 48109, USA}
\affiliation{University of Michigan, 500 S. State Street, Ann Arbor, MI 48109, USA}
\author[0000-0002-3461-0320]{Joseph Harry Silber}
\affiliation{Lawrence Berkeley National Laboratory, 1 Cyclotron Road, Berkeley, CA 94720, USA}
\author{David Sprayberry}
\affiliation{NSF NOIRLab, 950 N. Cherry Ave., Tucson, AZ 85719, USA}
\author[0000-0003-1704-0781]{Gregory Tarlé}
\affiliation{University of Michigan, 500 S. State Street, Ann Arbor, MI 48109, USA}
\author{Benjamin Alan Weaver}
\affiliation{NSF NOIRLab, 950 N. Cherry Ave., Tucson, AZ 85719, USA}
\author[0000-0001-5381-4372]{Rongpu Zhou}
\affiliation{Lawrence Berkeley National Laboratory, 1 Cyclotron Road, Berkeley, CA 94720, USA}
\author[0000-0002-6684-3997]{Hu Zou}
\affiliation{National Astronomical Observatories, Chinese Academy of Sciences, A20 Datun Road, Chaoyang District, Beijing, 100101, P.~R.~China}

\begin{abstract}

The Milky Way Survey of the Dark Energy Spectroscopic Instrument (DESI) has so far observed three classical dwarf spheroidal galaxies (dSphs): Draco, Sextans and Ursa Minor. Based on the observed line-of-sight velocities and metallicities of their member stars, we apply the axisymmetric Jeans Anisotropic Multi-Gaussian Expansion modeling (\textsc{jam}) approach to recover their inner dark matter distributions. In particular, both the traditional single-population Jeans model and the multiple population chemodynamical model are adopted. With the chemodynamical model, we divide member stars of each dSph into metal-rich and metal-poor populations. The metal-rich populations are more centrally concentrated and dynamically colder, featuring lower velocity dispersion profiles than the metal-poor populations. We find a diversity of the inner density slopes $\gamma$ of dark matter halos, with the best constraints by single-population or chemodynamical models consistent with each other. The inner density slopes are $0.71^{+0.34}_{-0.35}$, $0.26^{+0.22}_{-0.12}$ and $0.33^{+0.20}_{-0.16}$ for Draco, Sextans and Ursa Minor, respectively. We also present the measured astrophysical J and D factors of the three dSphs. Our results indicate that the study of the dark matter content of dSphs through stellar kinematics is still subject to uncertainties behind both the methodology and the observed data, through comparisons with previous measurements and data sets.

\end{abstract}


\section{Introduction}
\label{sec:intro}

The standard $\Lambda$ cold dark matter ($\Lambda$CDM) cosmological model has successfully predicted the distribution of large-scale structures in our Universe \citep[e.g.][]{Yang2004,Cole2005,Eisenstein2005,Henriques2012,Dawson2013,Han2015,Wang2016,Springel2018,DESI2024c}. However, there are discrepancies between predictions of the standard cosmological model and observations on small scales \citep[see][for a review]{Bullock2017}. One of the hotly debated issues is the core-cusp problem \citep[e.g.][]{Flores1994,Moore1994,Burkert1995,Salucci2001,deBlok2001,Gentile2004,Gilmore2007,Salucci2009,deBlok2010,Oh2011,Salucci2019,Salucci2022}. CDM simulations predict dark matter halos with inner slopes close to 1 (cusp). By contrast, observations of low surface brightness galaxies, gas-rich dwarfs and dwarf spheroidal galaxies (dSphs) sometimes indicate inner density slopes close to 0 (core).

Different scenarios have been proposed to understand the core-cusp issue. Explanations beyond $\Lambda$CDM turn to alternative models, such as self-interacting dark matter \citep[e.g.][]{Spergel2000,Rocha2013,Oman2015,Foot2015,Kaplinghat2016,Jiang2023} and fuzzy dark matter \citep[e.g.][]{Hu2000,Hui2017}. 
Within the CDM framework, stellar feedback seems to be the most promising, that repeated supernovae explosions move dark matter to more extended orbits and lower the central densities \citep[e.g.][]{Navarro1996,Read2005,Read2019,Freundlich2020a,Li2023,Boldrini2021}, with supporting evidences from  simulations \citep[e.g.][]{Mashchenko2008,Pontzen2012,Pontzen2014}. Observationally, \cite{Read2019} reported an anti-correlation between the inner dark matter densities of dwarf galaxies and their stellar-to-halo mass ratios, supporting the stellar feedback scenario \citep{DiCintio2014}. \cite{Hayashi2020} summarized that the relation between the inner dark matter density slopes of MW dSphs and their stellar-to-halo mass ratios generally agrees with recent hydrodynamical simulations \citep[e.g.][]{Tollet2016,Freundlich2020b,Lazar2020}.

However, the constraints on the inner dark matter densities for dSphs are still limited by statistical errors \citep{Zhu2016} and systematic uncertainties behind dynamical modeling \citep{Genina2018,Wang2022,Wang2023}. For example, \cite{Read2019}, \cite{Kaplinghat2019}, \cite{Hayashi2020} and \cite{Andrade2024} all measured the inner dark matter densities for eight classical dSphs in our Milky Way (MW), utilizing spherical/axisymmetric Jeans analysis, or the phase-space distribution function method, but part of their measurements are not fully consistent with each other. In fact, cuspy dSphs can be biased to be cored due to violations of model assumptions \citep{Genina2018,Wang2022}. Besides, binary orbital motions can inflate the overall velocity dispersion, whereas deflate the very central densities of dSphs in dynamical modeling \citep[e.g.][]{2022ApJ...939....3P}, though the change is not expected to be significant for classical dSphs \citep{Wang2023}.

The MW dSphs are close enough to have their individual member stars observed for dynamical modeling \citep{Battaglia2013R,Battaglia2022}. Recent observations revealed the co-existence of multiple stellar populations in these systems \citep[e.g.][]{Tolstoy2004,Battaglia2008,AE2012b,Pace2020,Arroyo-Polonio2024}. Many studies model the stellar kinematics of different populations simultaneously to constrain the inner dark matter distributions in several dSphs \citep{Battaglia2008,Walker2011,AE2012a,Agnello2012,Amorisco2013,Zhu2016,Strigari2017,Hayashi2018,Pace2020}. For example, \cite{Walker2011} estimate the slopes of mass profiles for Fornax and Sculptor defined by masses enclosed within half-number radii of two stellar populations, and found that their results are consistent with cored dark matter halos. 

In this study, we use the chemodynamical model developed in \cite{Zhu2016} and \cite{2016MNRAS.462.4001Z}, which extends the discrete axisymmetric Jeans Anisotropic Multi-Gaussian Expansion (\textsc{jam}) model \citep{Watkins2013}. Specifically, we adopt both the single-population \textsc{jam} model and a chemodynamical model in our analysis. Because the total mass within the half-number radius\footnote{The radius within which it includes half of the stars in the sample.} has been proven to be a sweet point \citep[e.g.][]{Wolf2010,Walker2011,Wang2015,2017MNRAS.472.4786G,Wang2022}, which is not sensitive to mild deviations from model assumptions, joint modeling of multiple populations has the advantage of providing multiply measured total masses within the half-number radii of different populations, hence may bring better constraints than the single-population model. 

We mainly use spectroscopic data for member stars in dSphs from the MW Survey (MWS) of the Dark Energy Spectroscopic Instrument (DESI). DESI is currently one of the leading multi-object spectrographs for wide-field surveys \citep{Levi2013,DESI2016a,DESI2016b,DESI2022,Silber2023,Schlafly2023,DESI2024a}. The Early Data Release \citep[EDR;][]{DESI2024b} and Data Release 1 (DR1) of its first year data \citep{2025arXiv250314745D} are publicly available. DESI primarily focuses on distant galaxies to explore the nature of dark energy and mapping the large-scale structure of our Universe \citep{DESI2024c,DESI2024d,DESI2024e,DESI2024f,DESI2024g,DESI2024h,2025arXiv250314738D}. DESI MWS also observes millions of stars within the MW and MW dSphs  \citep{Cooper2023}. The Year-3 data of DESI MWS includes member stars in Draco, Sextans and Ursa Minor.

This paper is organized as follows. We first introduce spectroscopic data from DESI MWS for three dSphs and member star selection in Section \ref{sec:data}. The single-population and chemodynamical model are described in Section~\ref{sec:method}. We present our results in Section~\ref{sec:results}, including measurements of the Astrophysical factors for dark matter detections. We compare with previous measurements in Section~\ref{sec:disc}, including discussions on the core-cusp problem and selection effects. Our conclusions are presented in Section~\ref{sec:concl}.

\begin{figure}[h]
    \centering
    \begin{minipage}{\linewidth}
        \centering
        \includegraphics[width=0.8\linewidth]{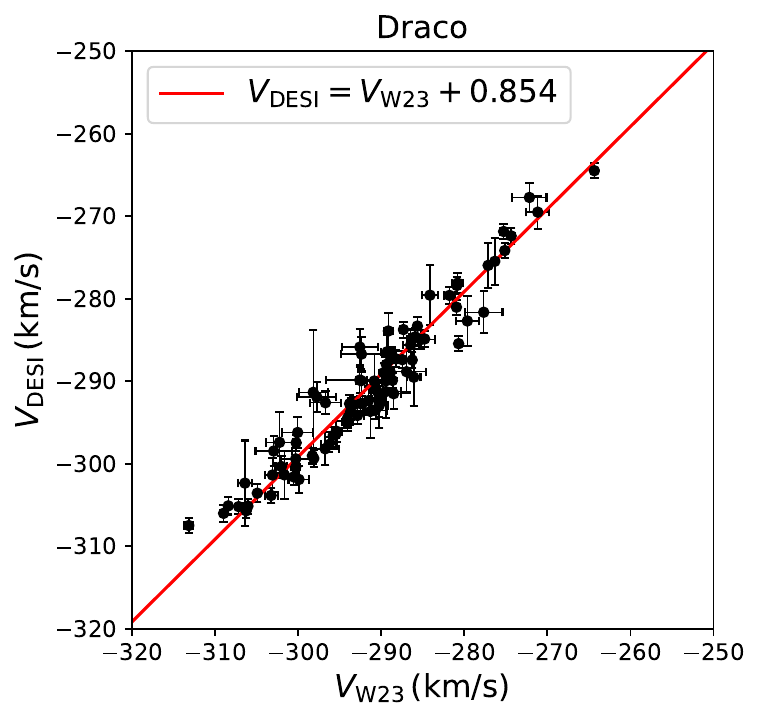}
    \end{minipage}
    \vfill
    \begin{minipage}{\linewidth}
        \centering
        \includegraphics[width=0.75\linewidth]{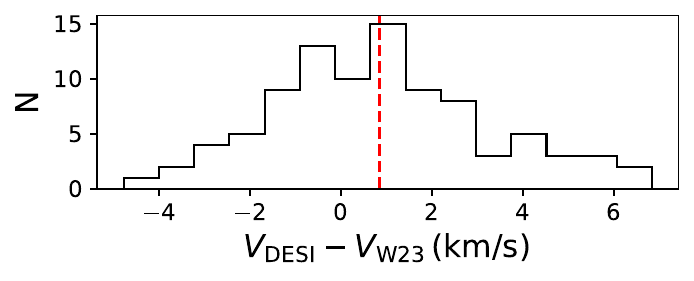}
    \end{minipage}
    \caption{{\bf Top:} A comparison between the Draco member star LOSV measurements by DESI MWS ($y$-axis) and \cite{Walker2023} ($x$-axis). The red line is the best fit with slope of 1, and its $y$-intercept is the offset between two observations (see the legend). {\bf Bottom:} The histogram of the differences between the Draco member star LOSV measurements by DESI MWS and \cite{Walker2023}. The red dashed vertical line represents the offset between two observations determined from the fit in the top panel.}
    \label{fig:draco_vlos}
\end{figure}

\begin{figure}[h]
    \centering
    \includegraphics[width=0.8\linewidth]{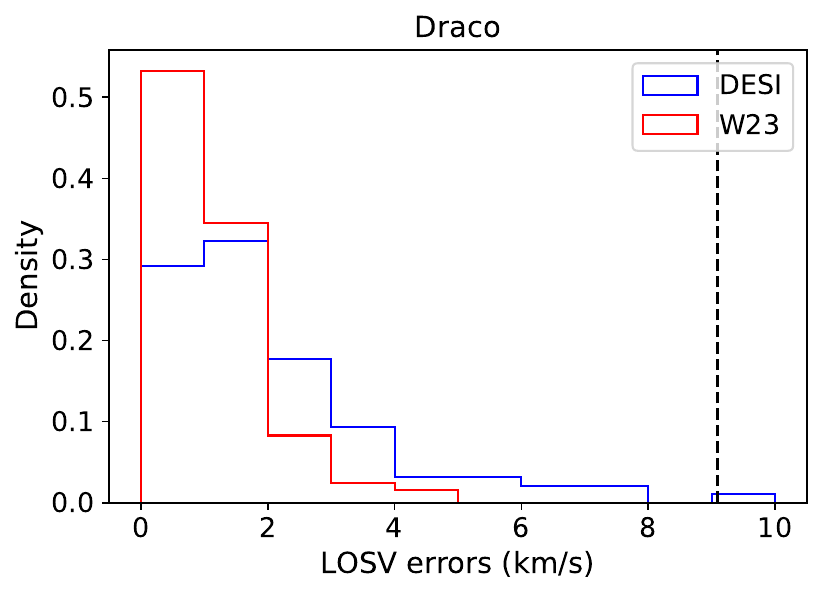}
    \caption{Distributions of LOSV errors for Draco member stars, observed by DESI MWS (blue) or by \cite{Walker2023} (red). The black vertical dashed line is the intrinsic LOSV dispersion of Draco (see Table~\ref{tab:dwarf prop}).}
    \label{fig:draco_vloserr}
\end{figure}

\section{Data}
\label{sec:data}

\subsection{The DESI Milky Way Survey}

The Dark Energy Spectroscopic Instrument (DESI) is currently one of the leading multi-object spectrographs for wide-field surveys \citep{DESI2016a,DESI2016b,DESI2022,DESI2024b}. The survey is carried on the Mayall 4 meter telescope at the Kitt Peak National Observatory \citep{Miller2024,Poppett2024}, with 5,000 fibers allocated over a $3.2^\circ$ diameter field of view at the prime focus. Although primarily focusing on observations of distant galaxies, DESI also collects spectroscopic data for stars. The DESI MWS mainly operates during bright time, which refers to nights with more significant moon light contamination \citep{Cooper2023,Koposov2024}. The main survey of MWS has observed nearly 12 million stars in the first three years. DESI MWS also has extra secondary targets, such as faint blue horizontal branch stars (BHBs). Beyond main survey and secondary targets, DESI MWS has tertiary programs, with dedicated fibers to observe special targets such as individual dSphs and stellar streams. In this work, we utilize the internal Year-3 data of DESI MWS. 

\subsection{DESI observations of Draco, Sextans and Ursa Minor}

So far, DESI MWS has observed stars in footprints of Draco, Sextans and Ursa Minor (UMi). The properties of them are summarized in Table \ref{tab:dwarf prop}. Nearly $85\%$ and $30\%$ of the data for Draco and Sextans are taken at Survey Validation (SV), with the remaining from the main survey after SV. The data for UMi are from the tertiary program. We use the line-of-sight velocity (LOSV) measurements in the heliocentric frame, and also metallicity measurements of observed stars output by the DESI MWS \texttt{RVS} pipeline. The readers can find more details in \cite{Cooper2023}, \cite{Koposov2024} and \cite{Koposov2025}. We select stars with $\texttt{RVS\_WARN}=0$ to ensure the robustness of the stellar model fit by the pipeline\footnote{$\texttt{RVS\_WARN}$ is a bitmask flag output by the \texttt{RVS} pipeline, which indicates potential issues or warnings in the LOSV measurements and stellar parameter determinations. The first bit of $\texttt{RVS\_WARN}$ is set to 1 if the difference in the $\chi^2$ values between the best-fit stellar model and the continuum-only model fit is small. The second bit is set to 1 if the LOSV is within $5 \, \mathrm{km/s}$ of the predefined velocity boundary ($ - 1500 $ to $1500 \, \mathrm{km/s}$). The third bit is set to 1, if the LOSV uncertainty exceeds $100 \, \mathrm{km/s}$. A spectrum without any of these concerns is indicated by $\texttt{RVS\_WARN}=0$}. For stars with repeated observations in different programs, we use the observation with the smallest error of the LOSV measurement. Note for the LOSV errors reported by  \texttt{RVS}, there is an additional systematic component or error floor of $0.9\,\mathrm{km/s}$, as mentioned in \cite{Cooper2023}. We add this systematic error in quadrature to the LOSV errors in our analysis.

\subsection{Observations from \cite{Walker2023}}
\label{subsec:combination}

We combine the DESI MWS data for three dSphs with the spectroscopic data from \cite{Walker2023}, which are based on Magellan/M2FS and MMT/Hectochelle observations. Since some stars have multi-epoch observations by \cite{Walker2023}, for each star we adopt the mean value of the LOSV (\text{vlos\_mean}) and metallicity (\text{feh\_mean}) measurements. The LOSV measurements by DESI and by \cite{Walker2023}, however, may subject to different zero points, and we use stars observed by both DESI and \cite{Walker2023} to correct for this systematic offset. We show the LOSVs of stars observed by both DESI and \cite{Walker2023} and in the footprint of Draco as an example in Figure \ref{fig:draco_vlos}. There is a good agreement between two sets of LOSVs \cite{Walker2023}. But the LOSV measured by \cite{Walker2023} of Draco is on average lower than that of DESI by about 1~km/s (see the histogram in the bottom panel). For each dSph, we determine the offset between two measurements as in Figure \ref{fig:draco_vlos}, and add this offset to the corresponding measurements of \cite{Walker2023}.

For stars existing in both DESI and \cite{Walker2023}, we use the LOSV measured with the smaller error. In Figure~\ref{fig:draco_vloserr}, we show the LOSV error distributions of stars in Draco, observed by DESI or in \cite{Walker2023}. In general, LOSV measurements from DESI have larger errors than \cite{Walker2023}, but both of them are smaller than the intrinsic LOSV dispersion of Draco (black dashed vertical line, from \cite{Pace2022}). For metallicity measurements, we also adopt the one with the smaller measurement error for stars existing in both DESI and \cite{Walker2023}, and we have corrected the offset of metallicity measurements by \cite{Walker2023} from those of DESI for each dSph.

\subsection{Member star selection}
\label{subsec:member star}

The membership probabilities for stars associated to known MW dwarf galaxies are calculated by \cite{Pace2022} according to their {\it Gaia} parallaxes and spatial distributions, with a projected Plummer model adopted as the prior for the spatial distribution. Specifically, we first select stars with the probability $\texttt{mem\_fixed}$ or $\texttt{mem\_fixed\_complete}$ greater than 0.9 from the membership catalog of \cite{Pace2022}. We cross match the sample of stars observed by DESI in the footprints of the three dSphs or in \cite{Walker2023} to these high probability member stars, to obtain our sample of spectroscopically observed member stars, with a searching radius of 0.3~arcsec. In the end, we further include $3\sigma$ clippings to the LOSVs and metallicities to remove stars at the tail of the distributions. For LOSV, we use the systemic LOSV and its dispersion in Table~\ref{tab:dwarf prop} to set the $3\sigma$ boundaries. For metallicity we use the mean and standard deviation of our samples. We also discard stars beyond $2.5$ half-light radius from the center of each dSphs to exclude possible interlopers and dwarf halo stars \citep[e.g.][]{Deason2022}.

The central coordinates and systemic velocities of each dSph have been subtracted from the observed coordinates and velocities for each member star in Cartesian coordinates, to correct the perspective rotation \citep{Feast1961}. The central coordinates and systemic proper motions are provided in Table~\ref{tab:dwarf prop}. The systemic LOSV is the mean LOSV of its member stars selected above. 

For each member star, we use their LOSVs and {\it Gaia} proper motions for dynamical modeling. However, at the distances of the three dSphs used in our analysis, the uncertainties in {\it Gaia} proper motion measurements are quite large. Nevertheless, we still incorporated proper motions in our dynamical modeling, with the observational uncertainties for both LOSVs and proper motions properly incorporated in the likelihood function (see Section~\ref{subsec:method single-pop} below). Notably, when comparing the observed and model-predicted velocity dispersions, we report the total velocity dispersions which include observational errors, instead of the intrinsic velocity dispersions deconvolved from them.

Though the pulsating velocities of DESI RR Lyrae stars have been corrected \citep{Medina2025a,Medina2025b}\footnote{The DESI MWS RRL targets are based on the {\it Gaia} Data Release 2 \citep[DR2;][]{Clementini2019} and the Pan-STARRS1 \citep[PS-1;][]{Sesar2017} catalogs.}, we want to eliminate the possibility that some RR Lyrae stars may still remain unidentified to inflate the velocity dispersions of our dSphs. We thus exclude all stars along the horizontal branch. We do not exclude possible binary stars since the effect of binary stars to inflate the velocity dispersion of classical dSph is minor \citep[e.g.][]{Hargreaves1996,Minor2010,Spencer2017,Wang2023}. Besides, member stars in UMi has multi-epoch observations in DESI (Qiu, et al., in preparation), based on which we have tried to exclude stars with large LOSV variabilities, but the change in LOSV dispersion is negligible. We provide in Table~\ref{tab:dwarf prop} the final number of member stars in the three dSphs after our selections above, with the number of stars observed by DESI or in \cite{Walker2023} provided as well.

\begin{table}
\centering
\begin{tabular}{lccc}
\hline
\rule{0pt}{2.8ex}\textbf{Property} & \textbf{Draco} & \textbf{Sextans} & \textbf{UMi} \\ \hline
$\alpha_0$ [deg] & 260.0684 & 153.2628 & 227.242 \\
$\delta_0$ [deg] & 57.9185 & $-1.6133$ & 67.2221 \\
$r_\mathrm{h}$ [arcmin] & $9.67 \pm 0.09$ & $16.50 \pm 0.10$ & $18.30 \pm 0.11$ \\
$\mu_{\alpha \star}$ [mas yr$^{-1}$] & 0.044 & $-0.409$ & $-0.120$ \\
$\mu_{\delta}$ [mas yr$^{-1}$] & $-0.188$ & 0.037 & 0.071 \\
$d$ [kpc] & 76 & 86 & 76 \\
$v_\mathrm{los}$ [km s$^{-1}$] & $-290.7$ & 224.3 & $-247.0$ \\
$\sigma_\mathrm{los}$ [km s$^{-1}$] & $9.1 \pm 1.2$ & $7.9 \pm 1.3$ & $8.6 \pm 0.3$ \\ \hline
$N_{\mathrm{tot}}$ & 407 & 440 & 1046 \\
$N_{\mathrm{DESI}}$ & 96 & 240 & 1003 \\
$N_{\mathrm{W23}}$ & 374 & 368 & 416 \\ \hline
\end{tabular}
\caption{A summary of the properties for Draco, Sextans and UMi. The top eight rows show the central sky coordinates in Equatorial coordinates ($\alpha_0$ and $\delta_0$), half-light radii ($r_h$), systemic proper motions ($\mu_\alpha\ast$ and $\mu_\delta$), distances ($d$), systemic LOSVs ($v_\mathrm{los}$), and intrinsic LOSV dispersions  ($\sigma_\mathrm{los}$). Distances are from \cite{Munoz2018}. All other data are from \cite{Pace2022}. The last three rows provide the numbers of member stars in each dSph. $N_{\mathrm{tot}}$ is the total number of member stars after selections in Section~\ref{subsec:member star}. $N_{\mathrm{DESI}}$ and $N_{\mathrm{W23}}$ are the numbers of member stars which are observed by DESI and from \cite{Walker2023}, respectively. There are overlaps between DESI and \cite{Walker2023}. For stars existing in both data set, we use the LOSV and metallicity measurements with smaller errors (see Section~\ref{subsec:combination}).}
\label{tab:dwarf prop}
\end{table}

\section{Method}
\label{sec:method}

We introduce both single-population and two-population chemodynamical models in this section.

\subsection{Single-population model}
\label{subsec:method single-pop}

Our single-population model is the axisymmetric Jeans Anisotropic Multi-Gaussian Expansion (\textsc{jam}) modeling method \citep{Cappellari2008,Watkins2013}. \textsc{jam} is an open source code\footnote{\url{https://github.com/lauralwatkins/cjam}}, and in this study we use a slightly different version by \cite{Zhu2016} and \cite{2016MNRAS.462.4001Z}, with improved python interfaces and plotting tools. 

Under the axisymmetric and steady-state assumptions, the Jeans equations in cylindrical coordinates are
\begin{align}
    \frac{\nu \left( \overline{v^2_R} - \overline{v^2_\phi}\right)}{R} + \frac{\partial \left( \nu \overline{v^2_R}\right)}{\partial R} + \frac{\partial \left(\nu \overline{v_R v_z} \right)}{\partial z} = -\nu \frac{\partial \Phi}{\partial R}, \label{eq:jeans eq1}\\
    \frac{\nu \overline{v_R v_z}}{R} + \frac{\partial \left(\nu \overline{v_R v_z} \right)}{\partial R} + \frac{\partial \left( \nu \overline{v^2_z}\right)}{\partial z} = -\nu \frac{\partial \Phi}{\partial z},
    \label{eq:jeans eq2}
\end{align}
where $\nu$ represents the tracer number density distribution and $\Phi$ represents the total gravitational potential. Throughout this paper, we use $(R, \phi, z)$ and $(x, y, z)$ to describe the cylindrical polar coordinate system and the Cartesian coordinate system in the intrinsic frame of the system, where $R^2=x^2+y^2$ and $z$ is the symmetry axis.

To obtain unique solutions to the Jeans equations, we further assume that the velocity ellipsoid is aligned with the cylindrical coordinates, so the cross-term can be neglected, i.e., $\overline{v_Rv_z}=0$. Besides, the velocity anisotropy, defined as $\beta_z=1-\overline{v^2_z}/\overline{v^2_R}$, is assumed as constant. Moreover, to calculate the first velocity moments, a rotation parameter $\kappa$ is defined as $\overline{v_\phi}=\kappa |\overline{v^2_\phi}-\overline{v^2_R} |^{1/2}$, which we also assume as constant. Our $\kappa$ follows the modified definition of \cite{Zhu2016}.

The gravitational potential ($\Phi$) in Equations \ref{eq:jeans eq1} and \ref{eq:jeans eq2} is contributed by both luminous and dark matter. For dark matter, we adopt the generalized NFW \citep{NFW1,NFW2,NFW3} (gNFW) model profile
\begin{equation}
    \rho(r)=\frac{\rho_s}{(r/r_s)^\gamma (1+r/r_s)^{(\alpha-\gamma)}},
\end{equation}
where $\rho_s$ is the scale density, $r_s$ is the scale radius, $\gamma$ and $\alpha$ are inner and outer density slopes, respectively. In our analysis we consider spherical dark matter halos. Although the spherical dark matter halo assumption may bias the inferred dark matter content \citep[e.g.][]{Pfenniger1994,Sylos2025}, we have checked that best constrained inner densities remain consistent well within $1\sigma$ if we consider the axisymmetric dark matter halo leaving its axis ratio as a free parameter. The best-fit axis ratios of dark matter halos for the three dSphs are slightly larger than 1 ($\sim1-1.5$). Moreover, because there are fewer member stars in outskirts, the outer density slope, $\alpha$, is weakly constrained. Thus we fix $\alpha$ to 4. We choose $4$ instead of $3$ because the outskirts of dSphs have undergone tidal effects so the slopes may be steeper \citep[e.g.,][]{Errani2021}, but we have verified that our constraints on the inner density profiles are not sensitive to the chosen outer slope, and the constraints on the inner density profiles do not change if $\alpha$ is a free parameter \citep[also see][]{Wang2022}. 

For the luminous matter, the gravitational potential is obtained by deprojecting the density map based on the inclination angle. The density map is estimated by multiplying the observed surface brightness map by the stellar mass-to-light ratio ($M_\ast/L_\ast$). Here $M_\ast/L_\ast$ is determined from the closest matched PARSEC isochrone \citep{parsec}\footnote{\url{http://stev.oapd.inaf.it/cmd}}. To obtain the surface brightness map, we create a projected image for each dSph using membership catalogs constructed in Section~\ref{subsec:member star}, with each member star weighted by its luminosity.  To account for faint stars with magnitudes below the \textit{Gaia} flux limit, we scale the surface brightness map up to bring the integrated luminosity equal to the total luminosity of each dSph from \cite{Munoz2018}. 

In addition to the dark and luminous potential, there is the tracer surface number density term ($\nu$). $\nu$ can be deprojected from the surface number density maps of member stars with spectroscopic observations, which are used as tracers. However, only a subset of stars in the full membership catalog constructed in Section~\ref{subsec:member star} have spectroscopic observations due to fiber collisions. Especially in central regions of dSphs, where the surface number density of stars is very high, the completeness fraction of spectroscopically observed stars versus targets is much lower \citep[e.g.][]{Ding2025}. Thus instead of using the surface number density maps of spectroscopically observed member stars, $\nu$ is deprojected from the surface number density maps of all target stars in the full membership catalog in Section~\ref{subsec:member star}. We do not include any scaling on the amplitude of $\nu$, as it cancels out on both sides of the Equations.

\begin{figure*}[htbp!]
    \centering
    \begin{minipage}{0.32\textwidth}
        \centering
        \includegraphics[width=\linewidth]{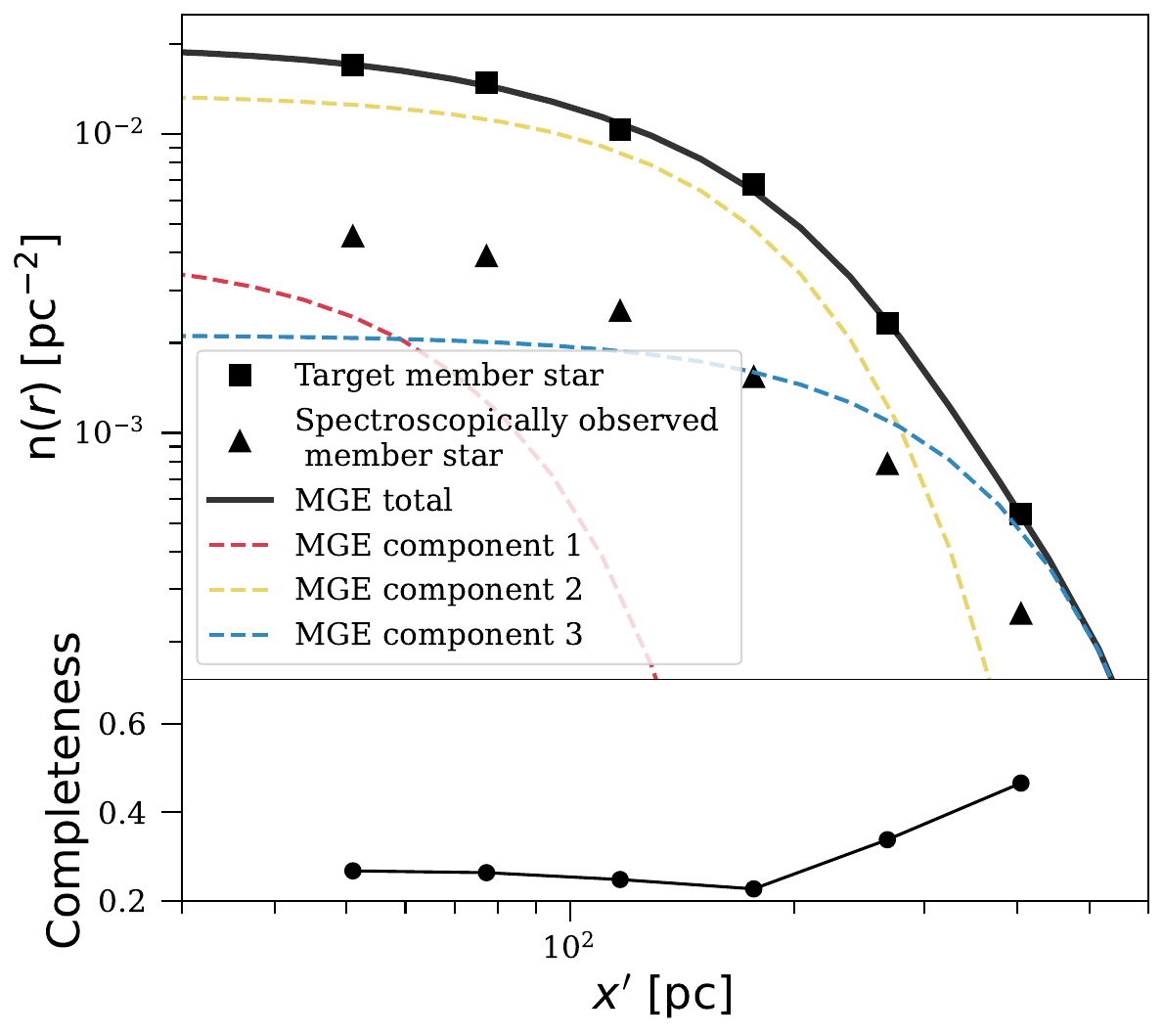}
        \text{(a) Draco}
    \end{minipage}
    \hfill
    \begin{minipage}{0.32\textwidth}
        \centering
        \includegraphics[width=\linewidth]{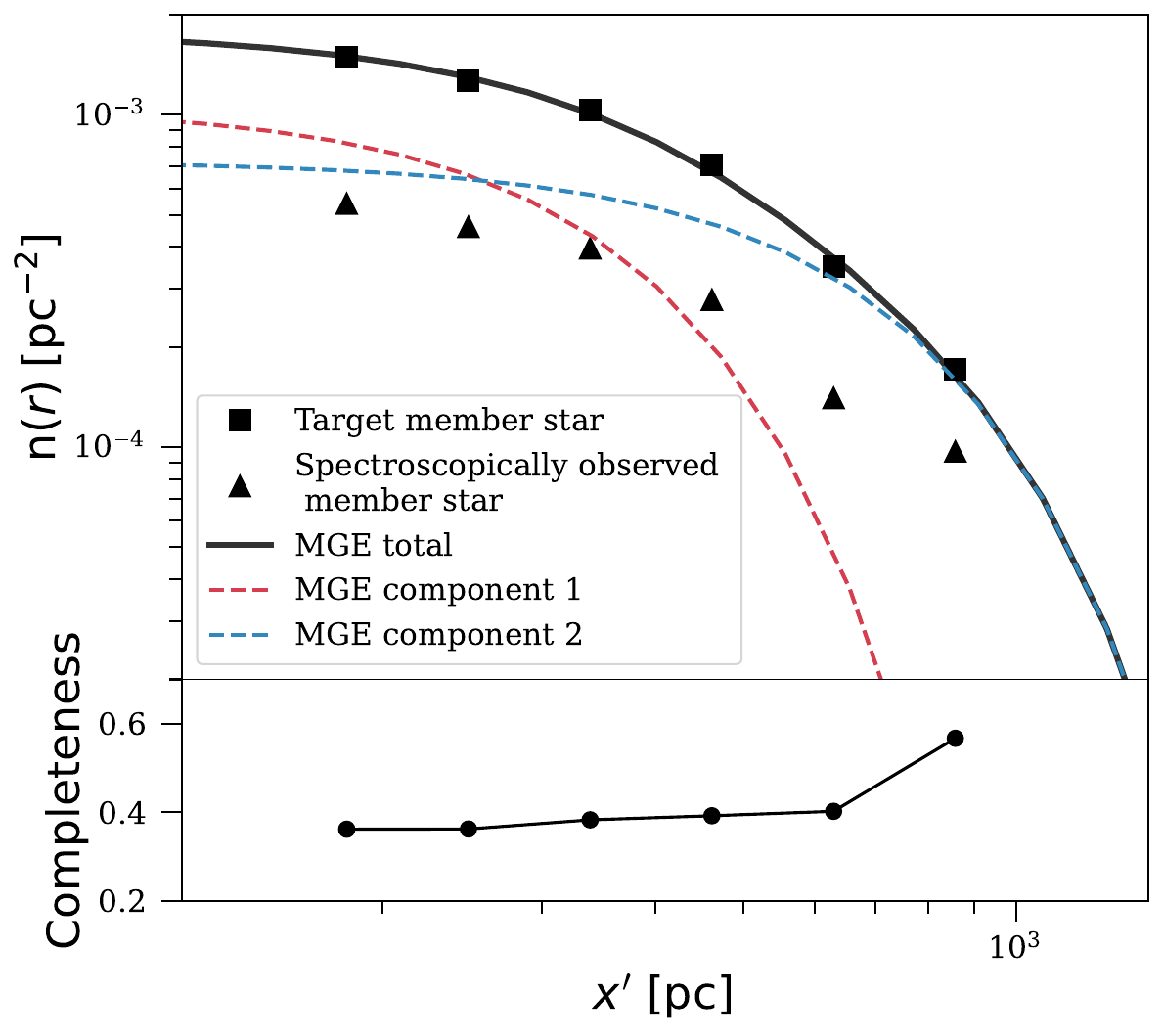}
        \text{(b) Sextans}
    \end{minipage}
    \hfill
    \begin{minipage}{0.32\textwidth}
        \centering
        \includegraphics[width=\linewidth]{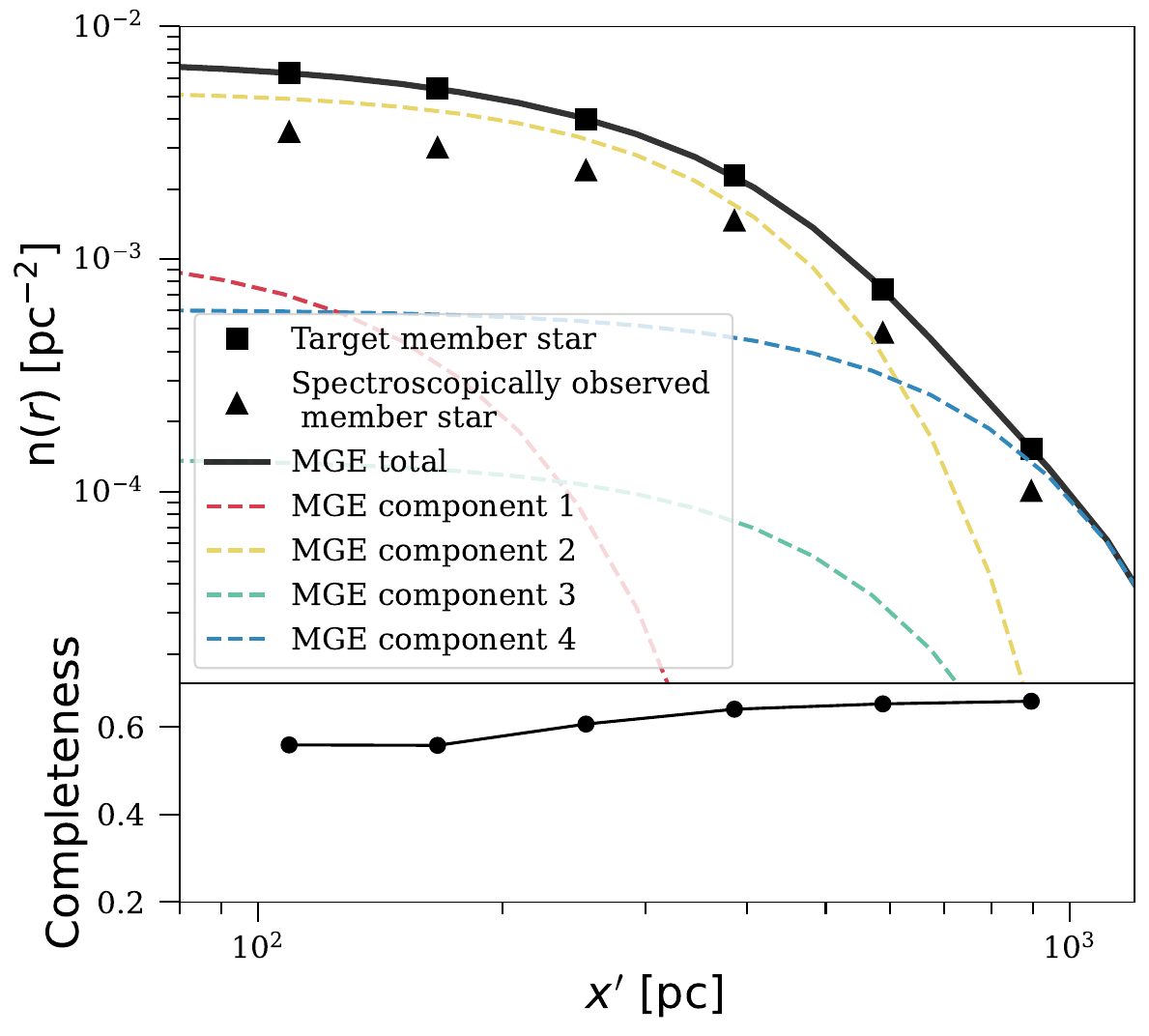}
        \text{(c) UMi}
    \end{minipage}
    \caption{{\bf Top panel:} Surface number density profiles of targets and spectroscopically observed member stars along the major axis, $x'$, for each dSph. The black squares and triangles are mean surface number densities in elliptical isophotal radial bins of targets and spectroscopically observed member stars. The colored dashed curves represent different MGE components, and the black solid curves are the summations of these MGE components, which match well the black squares. {\bf Bottom panel:} The completeness fraction of spectroscopically observed member stars versus targets, as a function of the projected distance along the major axis. The left, middle and right panels are for Draco, Sextans and UMi, respectively.}
    \label{fig:1sp surface number density}
\end{figure*}

We show the surface number density profiles of all target member stars (black squares) and spectroscopically observed member stars (black triangles) in concentric elliptical radial bins in Figure~\ref{fig:1sp surface number density}. The projected axis ratio $q'$ is obtained from the surface number density map and remains the same for different elliptical bins. The different dashed curves in the upper panels are the Multi-Gaussian Expansions, which will be introduced later. In the bottom panels, the completeness fractions for the spectroscopically observed member stars with respect to the targets in each dSphs are also shown. It can be seen that the completeness fractions are lower in the denser central regions due to unassigned fibers. Here the completeness fractions for Draco are different from those in \cite{Ding2025}, since we combine DESI MWS data with those from \cite{Walker2023} and remove stars on the horizontal branch.

Although the spatial distributions of spectroscopically observed member stars and the full member stars differ, we assume they do not differ in velocity moments, so the dynamical modeling outcome is not affected due to the incompleteness in fiber assignments. Equivalently, we are assuming that the spectroscopically observed member stars at different projected radii to the dSph center are a random subset in velocity space of the full sample. This is a reasonable assumption if the fiber assignment does not depend on any velocity information. However, as we have seen in Figure~\ref{fig:1sp surface number density}, the tiling and fiber completeness fractions differ in the central regions and the outskirts. In addition, metal-rich stars have more concentrated radial distributions to the dSph center, and the kinematics of metal-rich and metal-poor stars can differ. The fiber assignments may depend on member star kinematics. Unfortunately, if such a selection bias exists, it is hard to incorporate corrections, because we do not have LOSV measurements for the full target sample. We choose to adopt the assumption for the main results, with detailed discussions provided in Section~\ref{sec:disc} to discuss such selection effects.

In order to have quick analytical solutions to the Jeans equations, the Multi-Gaussian Expansion (MGE) approach is applied to decompose the surface number density maps of tracers, the surface density maps of luminous matter, and the gravitational potential of the dark matter profile into different Gaussian components. Each Gaussian component can be easily deprojected to get the corresponding analytical form in 3-dimensions. For example, the surface number density map of tracers $\Sigma(x', y')$ can be decomposed as
\begin{equation}
    \Sigma(x', y') = \sum_{j} \frac{L_j}{2\pi \sigma_j^2 q_j'} \exp \left[ -\frac{1}{2\sigma_j^2} \left( x'^2 + \frac{y'^2}{q_j'^2} \right) \right],
    \label{eq:2D MGE}
\end{equation}
where $L_j$, $\sigma_j$ and $q_j'$ are the total number density, dispersion and projected axis ratio of each Gaussian component $j$. Here $x'$ and $y'$ are the coordinates of the projected major and minor axes of the observed image of the galaxy. The corresponding 3-dimensional tracer number density distribution $\nu (R, z)$ (in cylindrical coordinates) can be simply deprojected as
\begin{equation}
    \nu (R, z) = \sum_{j} \frac{L_j}{(2\pi\sigma_j^2)^{3/2} q_j} \exp \left[ -\frac{1}{2\sigma_j^2} \left( R^2 + \frac{z^2}{q_j^2} \right) \right]
\end{equation}
for each Gaussian component. Here $q_j$ is the intrinsic axis ratio and can be linked to the projected axis ratio $q_j'$ through the inclination angle $i$ of the dSph
\begin{equation}
    q_j = \frac{\sqrt{q_j'^2 - \cos^2 i}}{\sin i}.
\end{equation}
The inclination angle $i$ is defined as the angle between the symmetry axis of the dSph and the line of sight, where $i = 0$ corresponds to a face-on orientation.

Since the surface number density map is noisy, in practice we fit 1-dimensional MGE to surface number density profiles in Figure~\ref{fig:1sp surface number density} to obtain $L_j$ and $\sigma_j$ for each Gaussian component $j$ in Equation~\ref{eq:2D MGE}. This is done in bins of concentric elliptical isophotes with the same axis ratio and orientation. Then these 1-dimensional MGE is extended to 2-dimensional MGE by adding the projected axis ratio $q_j'$ to each component. $q_j'$ is the axis ratio of the entire dwarf galaxy. Here we assume that different Gaussian components have the same axis ratio, which is reasonable according to the actual projected maps. We show the 1-dimensional MGE components for three dSphs in Figure~\ref{fig:1sp surface number density} as colored dashed curves, and the black solid curve denotes the summation of these MGE components. 
In all three dSphs, black solid curves provide good matches to black squares, which are surface number density profiles of targeted member stars.

The solution of the velocity moments is the summation of the analytical solution of each MGE component. In principle, the velocity anisotropy, $\beta_z$, and the rotation parameter, $\kappa$, can vary with different MGEs, but here we assume they are the same for all MGE components.

The solutions are first obtained in the intrinsic frames of the dSphs, and transformed to the observed frame. The coordinate system in the observed frame $(x^\prime,y^\prime,z^\prime)$ is a left-handed system. The $x^\prime$-axis lies along the projected major axis of the observed image of the galaxy, which is required by the MGE approach, and the $y^\prime$-axis lies along the projected minor axis. The $z^\prime$-axis is aligned with the line-of-sight direction, pointing away from us. On the other hand, the coordinate system in the intrinsic frame $(x,y,z)$ is a right-handed system. The $x$-axis coincides with the $x^\prime$-axis of the coordinates in the observed frame, and two coordinate systems are linked via
\begin{equation}
    \begin{pmatrix}
        x^\prime\\
        y^\prime\\
        z^\prime
    \end{pmatrix}
    =
    \begin{pmatrix}
        1 & 0        & 0      \\
        0 & -\cos{i} & \sin{i}\\
        0 & \sin{i}  & \cos{i}
    \end{pmatrix}
    \begin{pmatrix}
        x\\
        y\\
        z
    \end{pmatrix}
    ,
\end{equation}
where $i$ is the inclination angle of the galaxy.

We use a maximum-likelihood analysis to find the model that best describes our discrete data set. For a star located at $\boldsymbol{x}^\prime_i=(x^\prime_i,y^\prime_i)$ on the plane of the sky, it has the velocity vector $\boldsymbol{v}_i=(v_{x^\prime,i},\, v_{y^\prime,i},\, v_{z^\prime,i})$ and the error matrix
\begin{equation}
\boldsymbol{S}_i = 
    \begin{pmatrix}
    \sigma^2_{v_{x^\prime},i} & 0 & 0 \\
    0 & \sigma^2_{v_{y^\prime},i} & 0 \\
    0 & 0 & \sigma^2_{v_{z^\prime},i}
    \end{pmatrix}
    ,
\label{eq:error matrix}
\end{equation}
where $\sigma_{v_{x^\prime},i}$, $\sigma_{v_{y^\prime},i}$ and $\sigma_{v_{z^\prime},i}$ are observational errors of velocity components $v_{x^\prime,i}$, $v_{y^\prime,i}$ and $v_{z^\prime,i}$, respectively. Here the observational errors of $v_{x^\prime,i}$ and $v_{y^\prime,i}$ are obtained from their {\it Gaia} proper motion uncertainties. At the distances of the three classical dSphs, the proper motion uncertainties are very large, so the major constraints are contributed by the LOSVs. Including or not including {\it Gaia} proper motions for dSph member stars lead to almost the same results. Based on a set of parameters $\Theta$, the model predicts a mean velocity vector $\boldsymbol{\mu}_i=(\overline{u_{x^\prime,i}},\, \overline{u_{y^\prime,i}},\, \overline{u_{z^\prime,i}})$ with a covariance matrix
\begin{equation}
    \boldsymbol{C}_i = 
    \begin{pmatrix}
    \overline{u^2_{x^\prime}}-\overline{u_{x^\prime}}^2 & \overline{u^2_{x^\prime y^\prime}}-\overline{u_{x^\prime}}\overline{u_{y^\prime}} & \overline{u^2_{x^\prime z^\prime}}-\overline{u_{x^\prime}}\overline{u_{z^\prime}} \\
    \overline{u^2_{x^\prime y^\prime}}-\overline{u_{x^\prime}}\overline{u_{y^\prime}} & \overline{u^2_{y^\prime}}-\overline{u_{y^\prime}}^2 & \overline{u^2_{y^\prime z^\prime}}-\overline{u_{y^\prime}}\overline{u_{z^\prime}} \\
    \overline{u^2_{x^\prime z^\prime}}-\overline{u_{x^\prime}}\overline{u_{z^\prime}} & \overline{u^2_{y^\prime z^\prime}}-\overline{u_{y^\prime}}\overline{u_{z^\prime}} & \overline{u^2_{z^\prime}}-\overline{u_{z^\prime}}^2
    \end{pmatrix}
\label{eq:cov matrix}
\end{equation}
at the projected position of this star. By assuming that the velocity distribution predicted by the model is a trivariate Gaussian with the mean velocity $\boldsymbol{\mu}_i$ and the covariance $\boldsymbol{C}_i$, the likelihood for each star is
\begin{equation}
\begin{split}
    L_i &= p \left(\boldsymbol{v}_i \mid \boldsymbol{x^\prime}_i, \boldsymbol{S}_i, \boldsymbol{\mu}_i, \boldsymbol{C}_i\right) \\
    &= \frac{\exp{\left[-\frac{1}{2} \left(\boldsymbol{v}_i - \boldsymbol{\mu}_i  \right)^\mathrm{T} \left(\boldsymbol{C}_i + \boldsymbol{S}_i  \right)^{-1} \left(\boldsymbol{v}_i - \boldsymbol{\mu}_i  \right)\right]}}{\sqrt{(2\pi)^3 |\left(\boldsymbol{C}_i + \boldsymbol{S}_i  \right)|}}.
\end{split}
\end{equation}
The total likelihood is the product of $L_i$ for all $N$ stars
\begin{equation}
    L=\prod_{i=1}^N L_i~ .
\end{equation}

There are six free parameters in our single-population model, and their priors are:
\begin{enumerate}
    \item Scale density of the dark matter halo, $\rho_s$: uniform distribution over $-15 \leq \log_{10} [ \rho_s/(\mathrm{M_{\odot}} \, \mathrm{pc^{-3}}) ] \leq 15$;
    \item Scale radius of the dark matter halo, $r_s$: uniform distribution over $0.1 \leq  r_s/(\mathrm{kpc})  \leq 10$;
    \item Inner density slope of the dark matter halo, $\gamma$: uniform distribution over $0.1 \leq \gamma \leq 4$;
    \item Inclination angle of the galaxy, $i$: uniform distribution over $\cos^{-1} (q') < i/\text{deg} \leq 90$;
    \item Velocity anisotropy, $\beta_z$: uniform distribution over $-\infty < -\ln(1-\beta_z) < + \infty$;
    \item Rotation parameter, $\kappa$: uniform distribution over $-\infty < \kappa < + \infty$.
\end{enumerate}

To improve the efficiency of exploring the parameter space for $\rho_s$ and $r_s$, we fit redefined parameters $d_1=\log_{10}\rho_s^2r_s^3$ and $d_2=\log_{10}\rho_s$. Besides, the velocity anisotropy $\beta_z$ is symmetrically transformed as $\lambda=-\ln(1-\beta_z)$.

\subsection{Chemodynamical model} 

For our chemodynamical model, we follow the approach in \cite{Zhu2016,2016MNRAS.462.4001Z}. In general, for systems with $k$ different stellar populations, each population features distinct spatial, chemical and dynamical properties, but all trace the same gravitational potential. Specifically, here we consider $k=2$ stellar populations, composed of a metal-poor population and a metal-rich population. We consider chemical, spatial and dynamical probabilities for each star.

\subsubsection{Spatial probability}
\label{subsubsec:spatial probability}

For the star $i$ observed at $(x^\prime_i,y^\prime_i)$, the conditional probability of it belonging to population $k$ is defined as
\begin{equation}
    P_{\mathrm{spa},i}(k|x'_i,y'_i) = \frac{\Sigma^k (x^\prime_i,y^\prime_i)}{\Sigma_{\mathrm{tot}} (x^\prime_i,y^\prime_i)},
    \label{eq:spatialP}
\end{equation}
where the total surface number density $\Sigma_{\mathrm{tot}} (x^\prime,y^\prime)$ is the superposition of the surface number densities of two populations: $\Sigma_{\mathrm{tot}}=\Sigma^{\mathrm{metal-poor}}+\Sigma^{\mathrm{metal-rich}}$. 

For the spatial probability, if the maps of two populations are completely free, there are too many parameters which are difficult to constrain. To reduce the number of free parameters, we first construct two template maps, and the surface number density maps of two populations are assumed to be linear combinations of these two template maps.

To construct the template maps, we first apply a simple hard cut in metallicity to divide spectroscopically observed member stars into two template populations. This hard cut is chosen to maximize
the half-number radii difference between the two template populations after the division. Next we compute the surface number density maps of the two template populations defined by this hard cut. We call the two templates as red and blue. Note, however, the template maps created in this way are subject to incompleteness due to fiber collisions, and the completeness fraction is lower in more central regions where the surface number density of target member stars is higher. To correct for the incompleteness, we first calculate the completeness fraction map according to the ratios of the surface number density maps between spectroscopically observed member stars and the full sample of target stars in the membership catalog (see Section~\ref{subsec:member star}). We multiply the red and blue template maps by the inverse of the completeness fraction map to correct for the incompleteness. The completeness fraction map is the same for both templates, as it is impossible for us to divide targets into two populations.

After constructing the template maps, the surface number density map of each population in the chemodynamical model is the linear combination of these two template maps. For example, the surface number density map of the metal-poor population $\Sigma^{\mathrm{metal-poor}}$ can be written as
\begin{equation}
    \Sigma^{\mathrm{metal-poor}}(x^\prime,y^\prime)=h_1\Sigma^{\mathrm{red}}_{\mathrm{template}}(x^\prime,y^\prime) + h_2\Sigma^{\mathrm{blue}}_{\mathrm{template}}(x^\prime,y^\prime) \, ,
    \label{eq:template maps}
\end{equation}
where $\Sigma^{\mathrm{red}}_{\mathrm{template}}$ and $\Sigma^{\mathrm{blue}}_{\mathrm{template}}$ denote two template maps, and the fractions $h_1$ and $h_2$ will be left as free parameters. Since any two populations are divided from the total sample, we have $\Sigma_{\mathrm{tot}}=\Sigma^{\mathrm{metal-poor}}+\Sigma^{\mathrm{metal-rich}}=\Sigma^{\mathrm{blue}}_{\mathrm{template}}+\Sigma^{\mathrm{red}}_{\mathrm{template}}$. Hence, the surface number density map of the metal-rich population in the chemodynamical model has the same form but with different fractions $1-h_1$ and $1-h_2$. 

To create the red and blue template maps, a hard cut in metallicity is adopted above. In the end and with the best constrained chemodynamical model, we can calculate the probability for each star to be in the metal-rich or metal-poor populations, based on the joint posterior distribution combining spatial, kinematical and metallicity distributions (Equation~\ref{eq:relative probability} below). We call this a soft division. If the template maps are reasonable proxies for the maps of two populations, i.e., if the red/blue template is more metal-rich/metal-poor, the fractions are expected to be $h_1 \sim 0$ and $h_2 \sim 1$, and the hard and soft divisions would be close to each other.

\begin{figure*}[htbp!]
    \centering
    \begin{minipage}{0.32\textwidth}
        \centering
        \includegraphics[width=\linewidth]{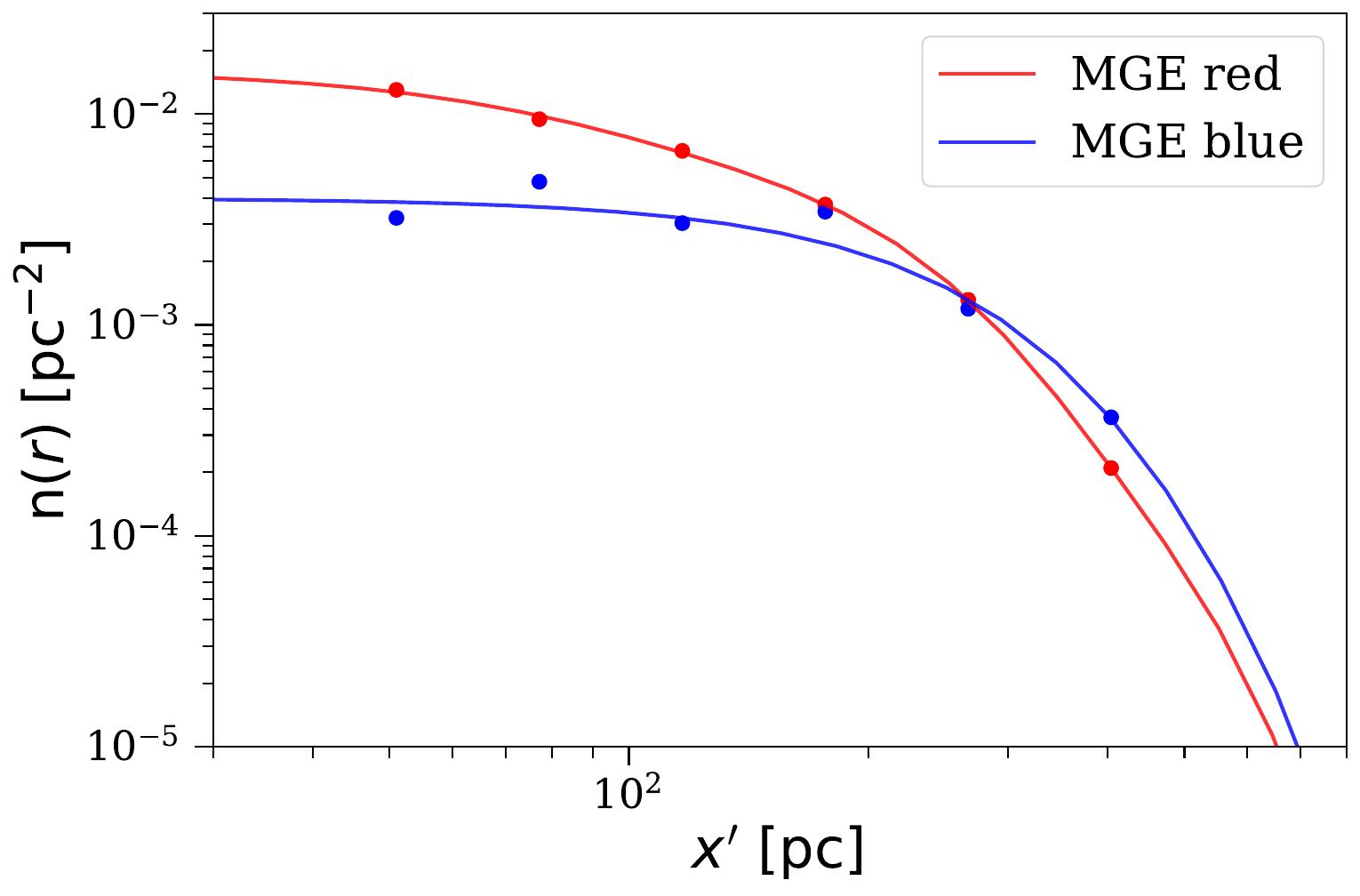}
        \text{(a) Draco}
    \end{minipage}
    \hfill
    \begin{minipage}{0.32\textwidth}
        \centering
        \includegraphics[width=\linewidth]{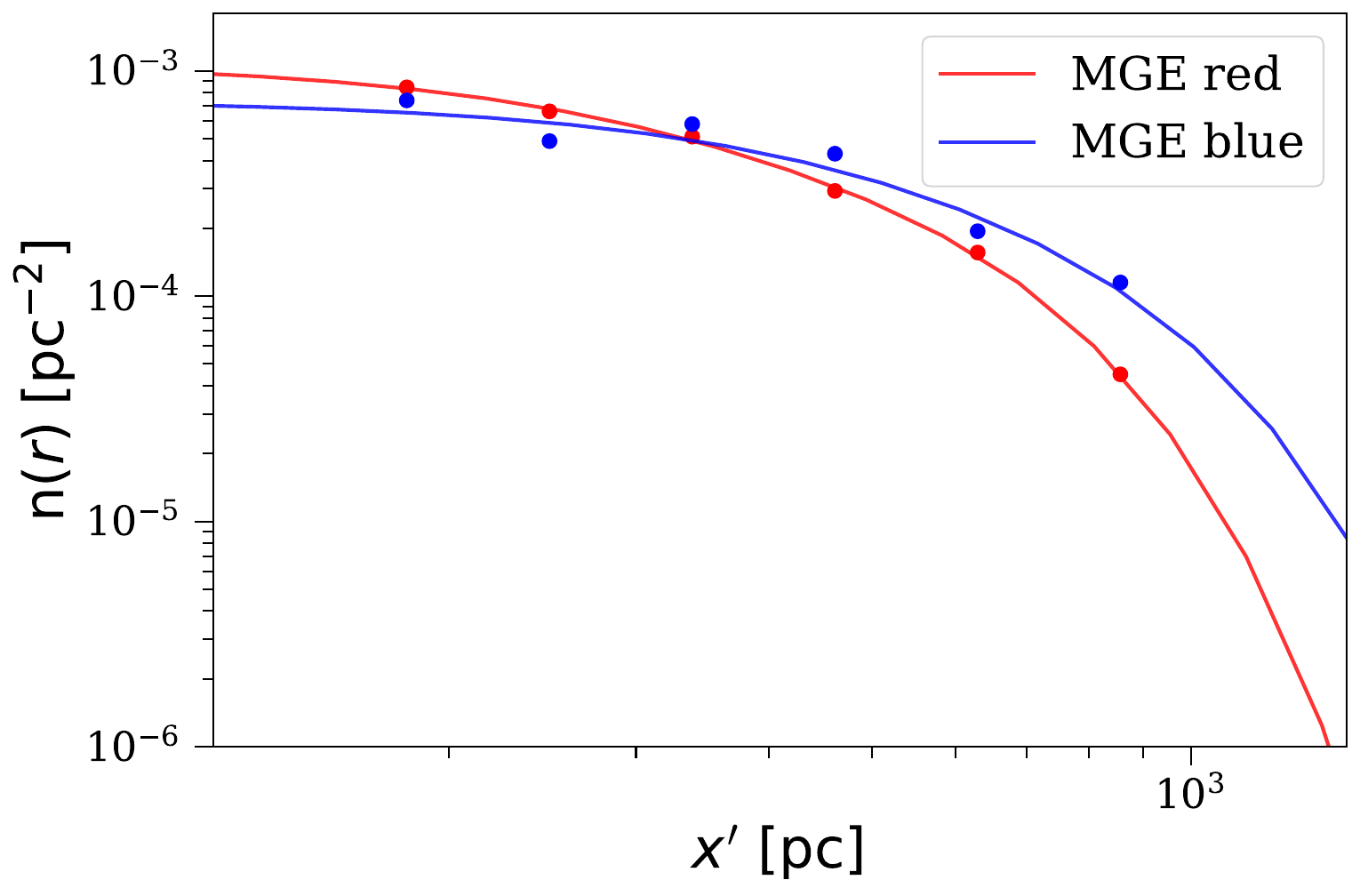}
        \text{(b) Sextans}
    \end{minipage}
    \hfill
    \begin{minipage}{0.32\textwidth}
        \centering
        \includegraphics[width=\linewidth]{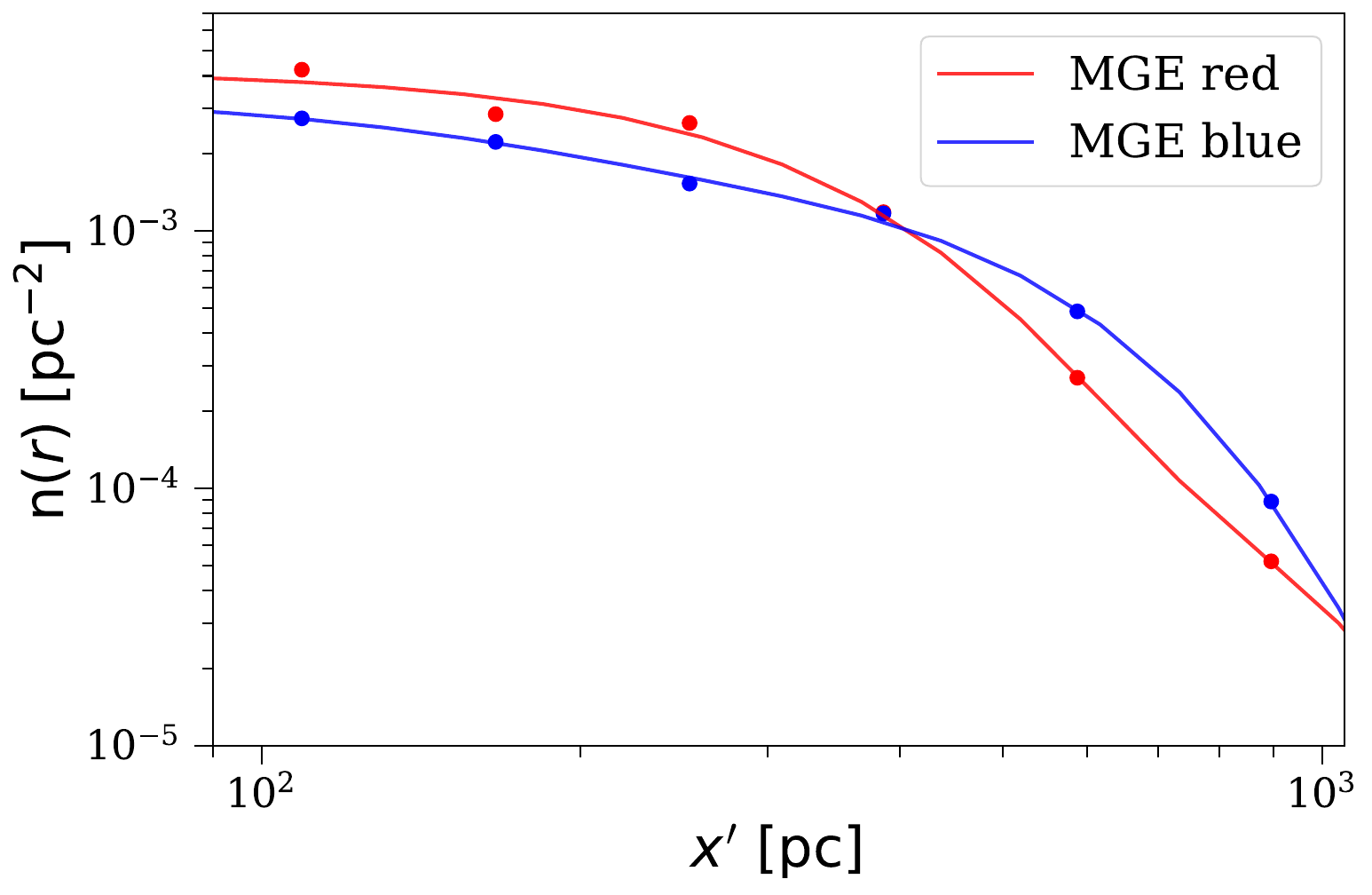}
        \text{(c) UMi}
    \end{minipage}
    \caption{Surface number density profiles of two template maps along the major axis, $x'$, for each dSph. The two template maps are obtained for the metal-rich (red) and metal-poor (blue) subpopulations after a hard cut in metallicity for the division, and the hard cut is chosen to maximize the difference in the half-number radii of the two populations. Red and blue dots are surface number densities directly calculated from corresponding template maps. Solid curves with the same colors as dots denote the reconstructed profiles from different MGEs. The left, middle and right panels are for Draco, Sextans and UMi, respectively.}
    \label{fig:2sp surface number density}
\end{figure*}

The template maps are shown in Figure \ref{fig:2sp surface number density}. The hard cuts in metallicity are $-2.10$, $-1.87$, $-2.07$ for Draco, Sextans and UMi, respectively. The red population has higher number densities in the inner region and lower densities in the outer region than the blue population. Thus, the hard cut in metallicity naturally leads to a more concentrated, metal-rich template population and a more extended, metal-poor template population.

\subsubsection{Chemical probability}

We assume that the metallicity distribution of each population $k$ follows a Gaussian distribution, characterized by the mean metallicity $Z_0^k$ and the dispersion $\sigma_Z^k$. Thus, for the star $i$ with the metallicity measurement $Z_i \pm \delta Z_i$, its chemical probability in the population $k$ can be written as

\begin{equation}
    P_{\mathrm{chm},i}(Z_i|k) = \frac{1}{\sqrt{2\pi [(\sigma_Z^k)^2 + (\delta Z_i)^2] }} \exp{\left[-\frac{1}{2}\frac{(Z_i-Z_0^k)^2}{[(\sigma_Z^k)^2 + (\delta Z_i)^2]}\right]}.
\end{equation}

\subsubsection{Dynamical probability}
For each population, the dynamical probability follows that in the single-population model. Each population featuring distinct number density distribution $\nu_k$ satisfies the Jeans Equations for their own velocity moments, but both of them trace the same gravitational potential. The number density distribution of each population is deprojected from its surface number density map as mentioned in Section \ref{subsubsec:spatial probability}. For the star $i$ with the velocity vector $\boldsymbol{v}_i=(v_{x^\prime,i},\, v_{y^\prime,i},\, v_{z^\prime,i})$ and the error matrix $\boldsymbol{S}_i$ (Equation~\ref{eq:error matrix} above), the dynamical probability of it being within the population $k$ is
\begin{equation}
    P_{\mathrm{dyn},i} (\boldsymbol{v}_i|k) = \frac{\exp{\left[-\frac{1}{2} \left(\boldsymbol{v}_i - \boldsymbol{\mu}_i^k  \right)^\mathrm{T} \left(\boldsymbol{C}_i^k + \boldsymbol{S}_i  \right)^{-1} \left(\boldsymbol{v}_i - \boldsymbol{\mu}_i^k  \right)\right]}}{\sqrt{(2\pi)^3 \left| \left(\boldsymbol{C}_i^k + \boldsymbol{S}_i  \right) \right|}},
\end{equation}
where $\boldsymbol{\mu}_i^k$ and $\boldsymbol{C}_i^k$ (Equation~\ref{eq:cov matrix} above) are the mean velocity and the covariance predicted at the position of the star by our model for each population, respectively. The velocity anisotropy $\beta_z^k$ and the rotation parameter $\kappa^k$ can vary between two populations, while they are still kept the same for all Gaussian components of each population.

\subsubsection{Joint probability}

For a star $i$ at projected coordinate $x'$ and $y'$, its joint probability distribution of population type $k$, velocity $\boldsymbol{v}_i$ and metallicity $Z_i$ is
\begin{equation}
    P(k,\boldsymbol{v}_i,Z_i|x',y')=P_{\mathrm{dyn},i}(\boldsymbol{v}_i|k,x',y')P_{\mathrm{chm},i}(Z_i|k)P_{\mathrm{spa},i}(k|x',y')~ .
    \label{eq:probability}
\end{equation}

To separate stars into different populations in a given model, we use the relative value
\begin{equation}
    P(k|\boldsymbol{v}_i,Z_i,x',y')=\frac{P(k,\boldsymbol{v}_i,Z_i|x',y')}{\sum_{k=1,2} P(k,\boldsymbol{v}_i,Z_i|x',y')},
    \label{eq:relative probability}
\end{equation}
to represent the probability for the star $i$ to be within the population $k$.

The likelihood of star $i$ can be written as
\begin{equation}
    L_i= \\
    \sum_{k=1,2} P_{\mathrm{dyn},i}(\boldsymbol{v}_i|k)P_{\mathrm{chm},i}(Z_i|k)P_{\mathrm{spa},i}(k|x',y'),
\end{equation}
where $\sum_{k=1,2} P_{\mathrm{spa},i}(k|x',y') =1 $ according to the definition of Equation~\ref{eq:spatialP} above.

Then the total likelihood for all $N$ stars is 
\begin{equation}
    L=\prod_{i=1}^N L_i~ .
\end{equation}

There are fourteen free parameters in our chemodynamical model, and their priors are:
\begin{enumerate}
    \item Scale density of the dark matter halo, $\rho_s$: uniform distribution over $-15 \leq \log_{10} [ \rho_s/(\mathrm{M_{\odot}} \, \mathrm{pc^{-3}}) ] \leq 15$;
    \item Scale radius of the dark matter halo, $r_s$: uniform distribution over $0.1 \leq  r_s/(\mathrm{kpc})  \leq 10$;
    \item Inner density slope of the dark matter halo, $\gamma$: uniform distribution over $0.1 \leq \gamma \leq 4$;
    \item Inclination angle of the galaxy, $i$: uniform distribution over $\cos^{-1} (q') < i/\text{deg} \leq 90$;
    \item Velocity anisotropies for the metal-rich population, $\beta_z^{\mathrm{mr}}$, and for the metal-poor population, $\beta_z^{\mathrm{mp}}$: uniform distribution over $-5 \leq -\ln(1-\beta_z^{\mathrm{mr}}) \leq +5$, $-5 \leq -\ln(1-\beta_z^{\mathrm{mp}}) \leq +5$;
    \item Rotation parameters for the metal-rich population, $\kappa^{\mathrm{mr}}$, and for the metal-poor population, $\kappa^{\mathrm{mp}}$: uniform distribution over $-\infty < \kappa^{\mathrm{mr}} < + \infty$, $-\infty < \kappa^{\mathrm{mp}} < + \infty$;
    \item Mean of metallicity distributions for the metal-rich population, $Z_0^{\mathrm{mr}}$, and for the metal-poor population, $Z_0^{\mathrm{mp}}$: uniform distribution over $-\infty < Z_0^{\mathrm{mr}} < + \infty$, $-\infty < Z_0^{\mathrm{mp}} < + \infty$;
    \item Dispersion of metallicity distributions for the metal-rich population, $\sigma_Z^{\mathrm{mr}}$, and for the metal-poor population, $\sigma_Z^{\mathrm{mp}}$: uniform distribution over $0 \leq \sigma_Z^{\mathrm{mr}} < + \infty$, $0 \leq \sigma_Z^{\mathrm{mp}} < + \infty$;
    \item Surface number density fractions for two template maps, $h_1$ and $h_2$: uniform distribution over $0 \leq h_1 \leq 1$, $0 \leq h_2 \leq 1$.
\end{enumerate}

Similar to the single-population model, we fit $d_1=\log_{10}\rho_s^2r_s^3$ and $d_2=\log_{10}\rho_s$ instead of $\rho_s$ and $r_s$. Besides, velocity anisotropies $\beta_z^{\mathrm{mr}}$ and $\beta_z^{\mathrm{mp}}$ are transformed as $\lambda^{\mathrm{mr}}=-\ln(1-\beta_z^{\mathrm{mr}})$ and $\lambda^{\mathrm{mp}}=-\ln(1-\beta_z^{\mathrm{mp}})$ as well.

\section{Results}
\label{sec:results}

In this section, we present our results for both single-population and chemodynamical models. For each model, we show best-fit model parameters as well as best-fit velocity moments compared with the real observations. For the chemodynamical model, we also present the spatial, kinematical and metallicity distributions of the two populations separated by the best model. In the end, we calculate the astrophysical J and D factors for indirect dark matter detections.

\subsection{Best constraints and performance of the single-population model}

\begin{figure*}[htbp!]
    \centering
    \begin{minipage}{\textwidth}
        \centering
        \includegraphics[width=0.6\textwidth]{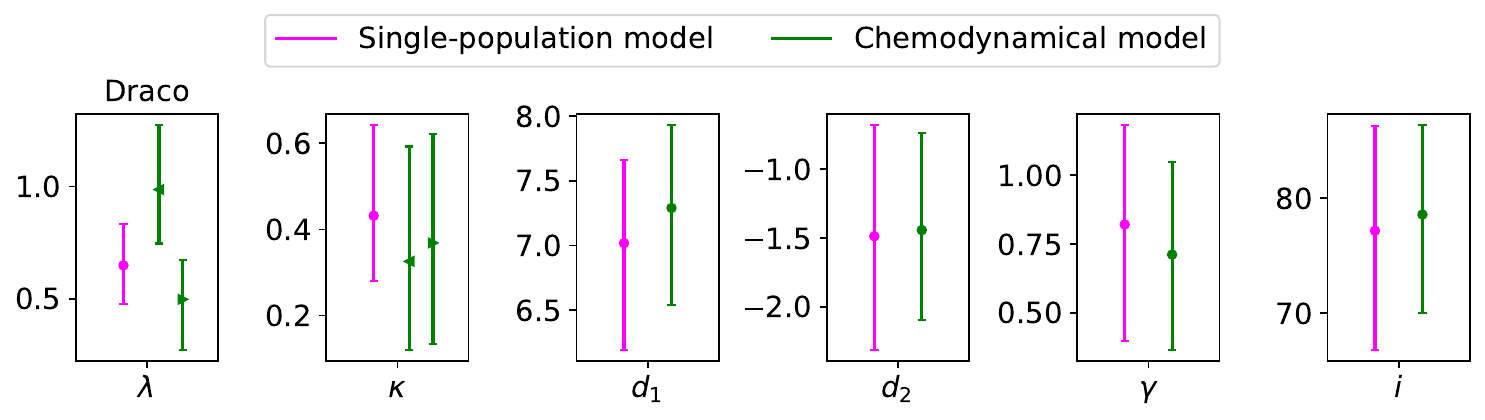}
    \end{minipage}
    \vfill
    \begin{minipage}{\textwidth}
        \centering
        \includegraphics[width=0.6\textwidth]{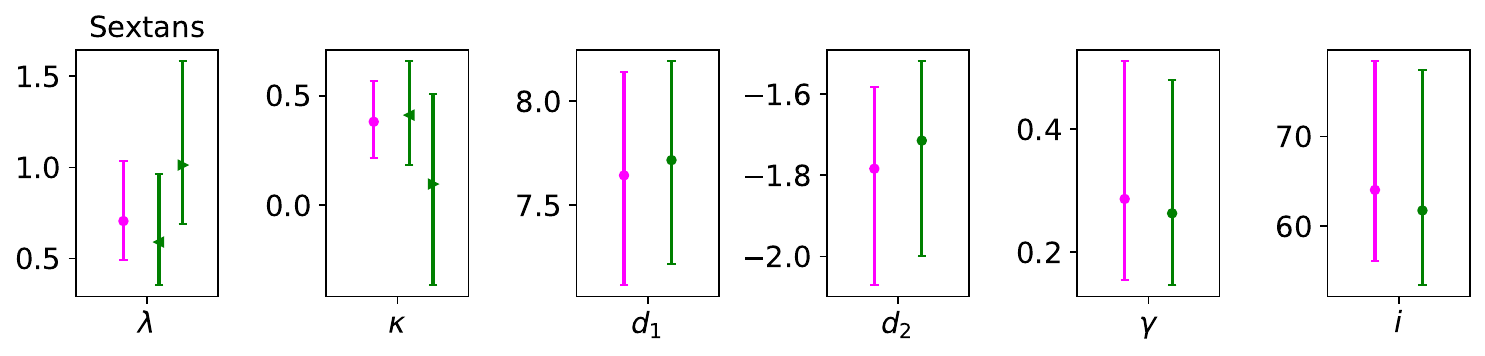}
    \end{minipage}
    \vfill
    \begin{minipage}{\textwidth}
        \centering
        \includegraphics[width=0.6\textwidth]{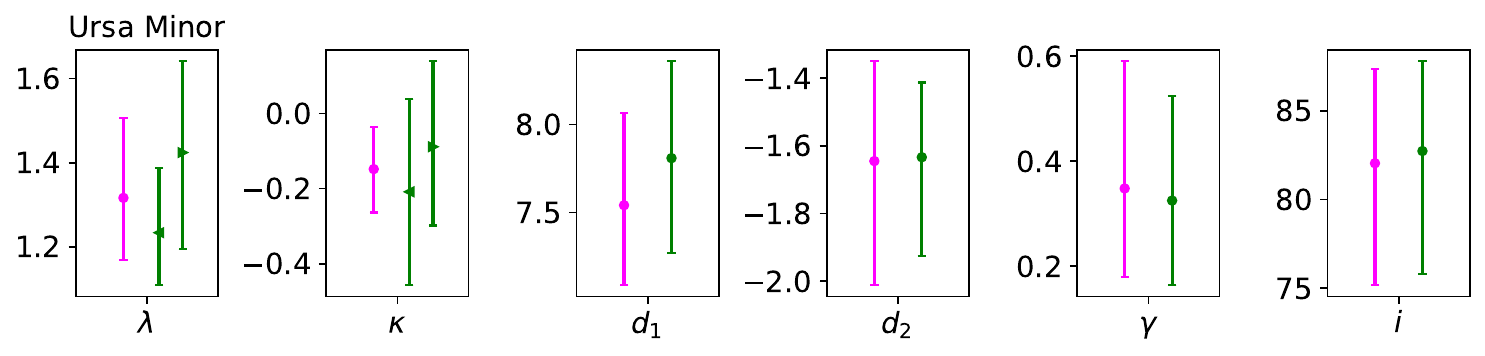}
    \end{minipage}
    \caption{Best-fit model parameters for the single-population (magenta) and chemodynamical (green) models, with the three rows refer to Draco, Sextans and UMi. The left and right green arrows in the first two columns for parameters $\lambda$ and $\kappa$ represent results for the metal-poor and metal-rich populations separated by the chemodynamical model. The $y$-axis ranges for different parameters are not the same. The meanings of different model parameters can be found in Section~\ref{sec:method}.}
    \label{fig:para compare}
\end{figure*}

The best-fit parameters for the single-population model of the three dSphs are provided in Table \ref{tab:parameters1}. In Figure \ref{fig:para compare}, we show these constraints for the single-population model by the magenta dots with errorbars. The green symbols denote those for the chemodynamical model, which will be discussed later in Section \ref{subsec:para in chemodyn}. 

\begin{table}[htbp]
\renewcommand{\arraystretch}{1.2}
\centering
\begin{tabular}{lccc}
\hline
\rule{0pt}{2.8ex}\textbf{Parameter} & \textbf{Draco} & \textbf{Sextans} & \textbf{UMi} \\
\hline
$\rho_s\ [\mathrm{M_{\odot}\,pc^{-3}}]$ 
 & $0.033^{+0.176}_{-0.028}$ 
 & $0.016^{+0.010}_{-0.008}$ 
 & $0.023^{+0.022}_{-0.013}$ \\
$r_s\ [\mathrm{kpc}]$ 
 & $2.12^{+3.87}_{-1.35}$ 
 & $6.22^{+2.50}_{-2.64}$ 
 & $4.70^{+3.17}_{-2.32}$ \\
$\gamma$ 
 & $0.82^{+0.36}_{-0.42}$ 
 & $0.29^{+0.23}_{-0.13}$ 
 & $0.35^{+0.24}_{-0.17}$ \\
$i\ [\mathrm{deg}]$ 
 & $77^{+9}_{-10}$ 
 & $64^{+14}_{-8}$ 
 & $82^{+5}_{-7}$ \\
$\lambda$ 
 & $0.65^{+0.19}_{-0.17}$ 
 & $0.71^{+0.33}_{-0.22}$ 
 & $1.32^{+0.19}_{-0.15}$ \\
$\kappa$ 
 & $0.43^{+0.21}_{-0.15}$ 
 & $0.38^{+0.19}_{-0.16}$ 
 & $-0.15^{+0.11}_{-0.12}$ \\
\hline
\end{tabular}
\caption{Best constrained model parameters based on the single-population model of three dSphs. The best-fit values are the medians of MCMC post-burn distributions, and errors are the 16th and 84th percentiles. The meanings of different model parameters can be found in Section~\ref{sec:method}.}
\label{tab:parameters1}
\end{table}

\begin{figure}[h]
    \centering
    \includegraphics[width=0.49\textwidth]{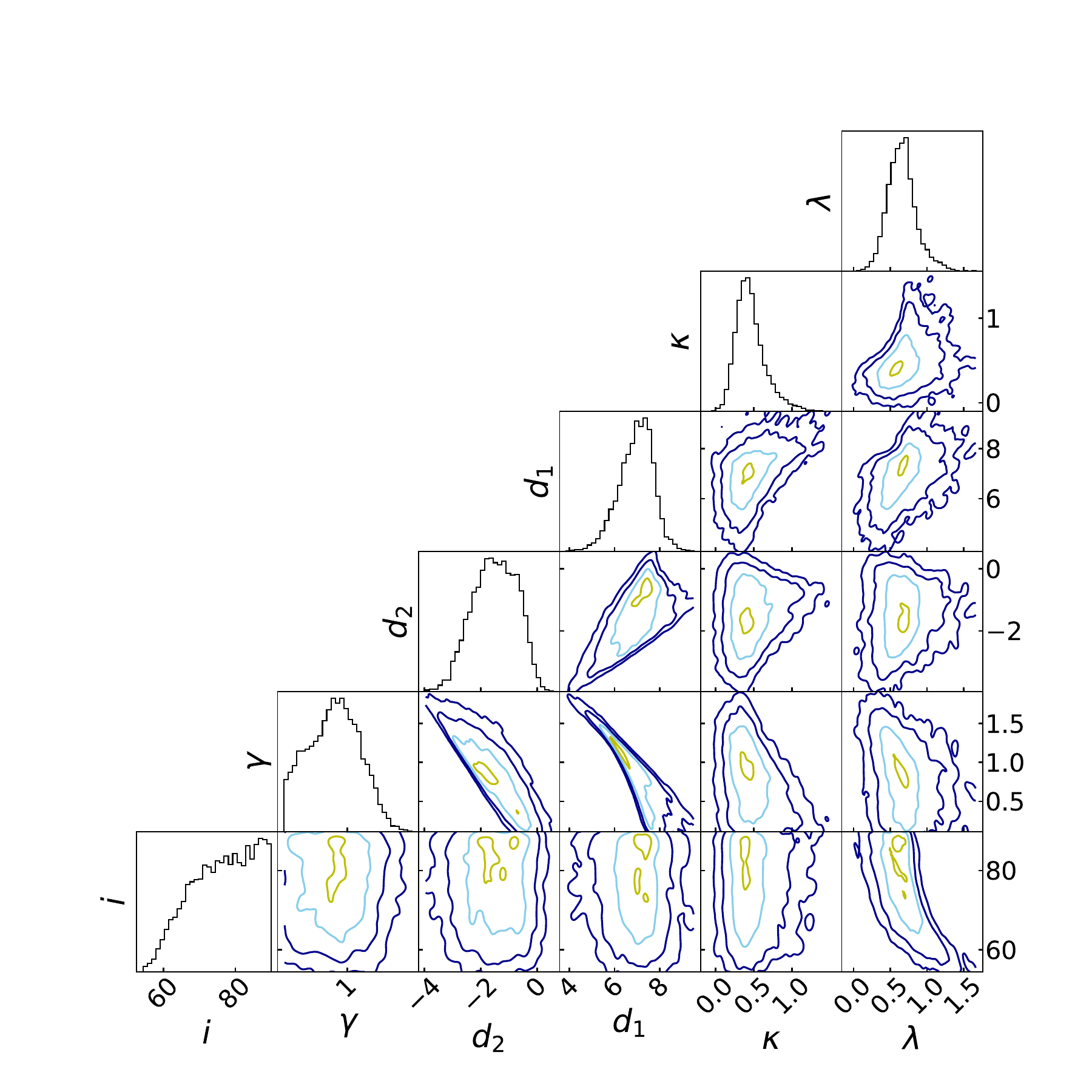}
    \caption{The likelihood contours of six model parameters for Draco. The meanings of different model parameters can be found in Section~\ref{sec:method}. The yellow, light blue and two dark blue contours represent the $10\%$, $1\sigma$, $2\sigma$ and $3\sigma$ regions of the MCMC post-burn distributions, respectively.}
    \label{fig:contour1 draco}
\end{figure}

The likelihood contours for different parameter combinations in the single-population model are shown in Figure \ref{fig:contour1 draco} for Draco only. The trends are similar for Sextans and UMi. Our model imposes weaker constraints on the inclination angle ($i$) compared to other parameters, which is indicated by its wide posterior distribution. The rotation parameter ($\kappa$) shows a weak correlation with the potential parameters ($d_1$, $d_2$ and $\gamma$), whereas there are strong degeneracies among the potential parameters themselves. The correlation between the velocity anisotropy ($\lambda$) and the other potential parameters is also present, which is the mass-anisotropy degeneracy. 

The mass-anisotropy degeneracy can be weakened if we have precise proper motion measurements. However, at the distances of the three dSphs used in this paper, their {\it Gaia} proper motions used in our constraints are subject to very large errors, and as we have mentioned, including or not including {\it Gaia} proper motions in our analysis leads to very similar constraints. More precise proper motions might be achieved by combining {\it Gaia} and previous observations by the Hubble Space Telescope \citep[HST, e.g.][]{delPino2022}, but this is only available for a small number of stars in the center. Thus in this study we only use {\it Gaia} proper motions. In fact, \cite{Vitral2024} recently combined precise proper motion measurements for about 400 stars in Draco from HST observations with LOSV measurements from \cite{Walker2023}, and performed modeling similar to \textsc{jam}. Their inferred inner dark matter density slope is fully consistent with our result.

The best recovered dark matter density profiles for the single-population model are shown by the black solid curves in Figure~\ref{fig:density profile}, which we will discuss slightly later in Section~\ref{subsec:para in chemodyn} below, together with the chemodynamical model.

\begin{figure*}[htbp!]
    \centering
    \begin{minipage}{0.9\textwidth}
        \centering
        \includegraphics[width=\linewidth]{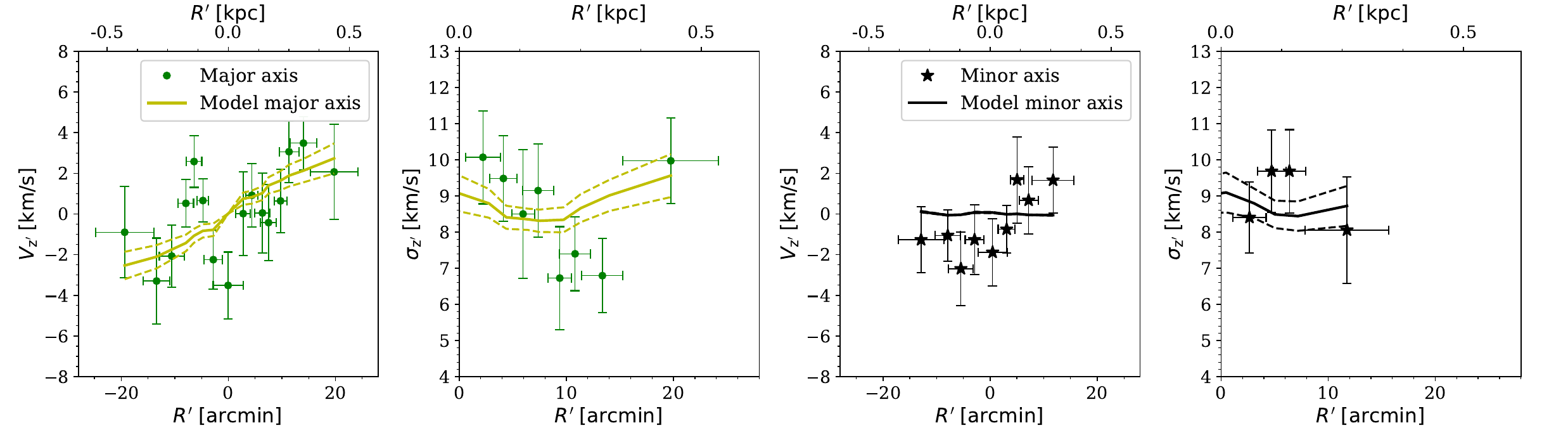}
        \text{(a) Draco}
    \end{minipage}
    \vfill
    \begin{minipage}{0.9\textwidth}
        \centering
        \includegraphics[width=\linewidth]{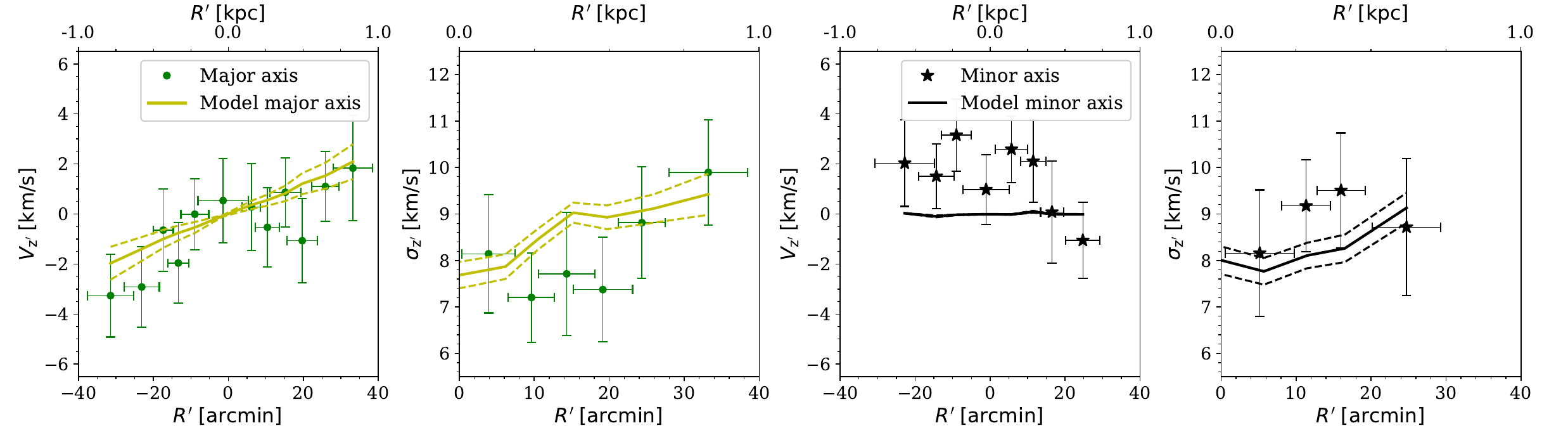}
        \text{(b) Sextans}
    \end{minipage}
    \vfill
    \begin{minipage}{0.9\textwidth}
        \centering
        \includegraphics[width=\linewidth]{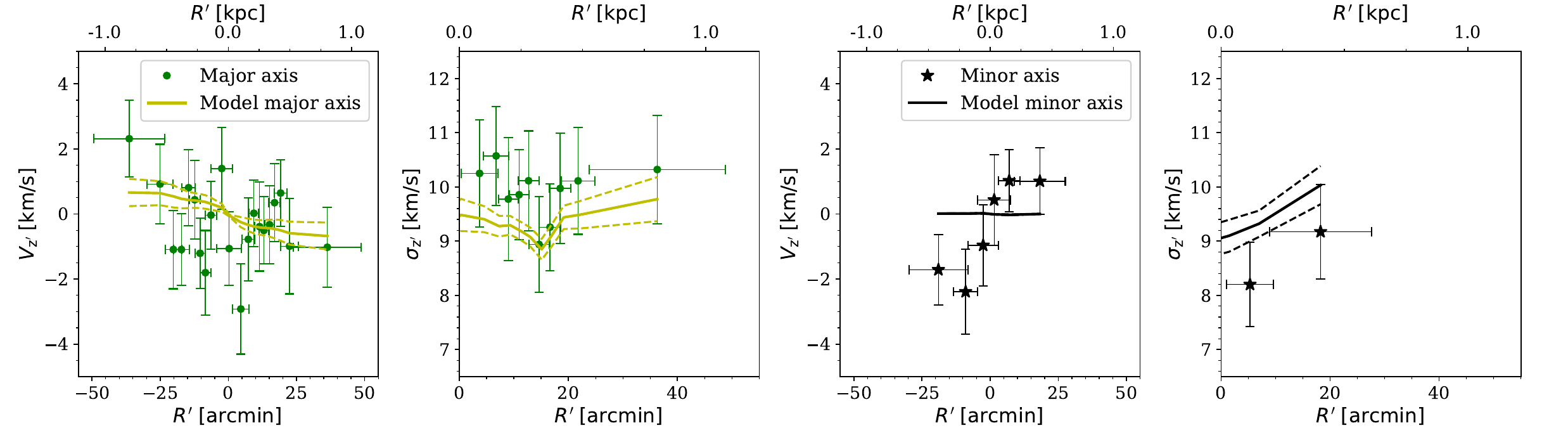}
        \text{(c) UMi}
    \end{minipage}
    \caption{The mean LOSV ($v_{z^\prime}$) and the LOSV dispersion ($\sigma_{z'}$) profiles along major (two left columns) and minor (two right columns) axes of three dSphs: (a) Draco, (b) Sextans and (c) UMi. The green dots and black stars represent the binned data within sectors of $\pm45^\circ$ relative to the corresponding axes. The number of stars in each bin is the same for a given galaxy, but varies between different systems. The $x$ errors represent the bin width while the $y$ errors represent the 1$\sigma$ scatters. The yellow and black curves represent best-fit model profiles along the projected major and minor axes, respectively. Solid and dashed curves represent best-fit model profiles and corresponding 1$\sigma$ uncertainties. The velocity dispersions of both the data and the best-fit model include observational uncertainties. They are not intrinsic velocity dispersions deconvolved with errors.}
    \label{fig:vz_1sp}
\end{figure*}

The observed and best-fit first (mean velocity) and second (velocity dispersion) velocity moments are shown in Figure \ref{fig:vz_1sp}. The mean velocity and velocity dispersion profiles are calculated from the LOSVs along $z^\prime$. In general, the velocity moments predicted by the best models are consistent with data, with the velocity dispersion profiles show good matches in amplitudes. The model can also recover the general shapes in the velocity dispersion profiles. In the velocity dispersion profile along the major axis of Draco and Sextans (second panels from the left in the top and middle rows), however, a few green dots drop below the model curves at $R'\sim10-20 \ \mathrm{arcmin}$. This is likely driven by the outer most green dots, that causes the model to match the velocity dispersion in the very outer region, sacrificing the match at $R'\sim10-20 \ \mathrm{arcmin}$.

For all three dSphs, rotations around major axes (green dots in the leftmost panels) are present and can be produced by the model (yellow curves). However, the matches between the black stars and the model predicted black curves (along minor axis) are poor. For axisymmetric model, the rotation is expected to be revealed in the LOSVs along the projected major axis, and this is the reason why the model can only predict a flat black curve in right-center panels along the minor axis. The deviations of black stars from zero may reflect the misalignment between minor and rotation axes in the real systems, which indicates these systems are deviating from axis-symmetry, as have been seen in simulated dSphs \citep{Kowalczyk2018}.

\subsection{Best constraints and performance of the chemodynamical model}
\label{subsec:para in chemodyn}

The full list of best-fit parameters for the chemodynamical model is shown in Table \ref{tab:parameters2}. For the fractions $h_1$ and $h_2$ of metal-poor stars in red and blue templates (see Equation \ref{eq:template maps}), they are close to 0 and 1, respectively. This shows that our template maps, though created through simpler hard cuts in stellar metallicities, lead to similar divisions in the stellar populations as the soft divisions according to posterior distributions of the best constrained model, which takes into account the joint spatial, kinematical and metallicity probabilities.

We also present the best constrained parameters for the chemodynamical model by green symbols in Figure \ref{fig:para compare}. The best-fit parameters for the single-population and chemodynamical models are consistent with each other. For the tracer parameters ($\lambda$ and $\kappa$), constraints of two populations in the chemodynamical model feature larger errorbars than those of the single-population model. This is because the sample size of each population is smaller than that of the total sample.

\begin{table}[htbp]
\renewcommand{\arraystretch}{1.2}
\centering
\begin{tabular}{lccc}
\hline
\rule{0pt}{2.8ex}\textbf{Parameter} & \textbf{Draco} & \textbf{Sextans} & \textbf{UMi} \\
\hline
$\rho_s\ [\mathrm{M_{\odot}pc^{-3}}]$ & $0.036^{+0.147}_{-0.028}$ & $0.019^{+0.011}_{-0.009}$ & $0.023^{+0.015}_{-0.011}$ \\
$r_s\ [\mathrm{kpc}]$                 & $2.58^{+3.52}_{-1.67}$    & $5.82^{+2.71}_{-2.39}$    & $5.62^{+2.89}_{-2.53}$    \\
$\gamma$                              & $0.71^{+0.34}_{-0.35}$    & $0.26^{+0.22}_{-0.12}$    & $0.33^{+0.20}_{-0.16}$    \\
$i\ [\mathrm{deg}]$                   & $79^{+8}_{-9}$            & $62^{+12}_{-8}$           & $83^{+5}_{-7}$            \\
$\lambda^{\mathrm{mr}}$               & $0.50^{+0.17}_{-0.23}$    & $1.01^{+0.57}_{-0.32}$    & $1.42^{+0.22}_{-0.23}$    \\
$\lambda^{\mathrm{mp}}$               & $0.99^{+0.29}_{-0.24}$    & $0.59^{+0.38}_{-0.24}$    & $1.23^{+0.16}_{-0.13}$    \\
$\kappa^{\mathrm{mr}}$                & $0.37^{+0.25}_{-0.24}$    & $0.10^{+0.41}_{-0.46}$    & $-0.09^{+0.23}_{-0.21}$   \\
$\kappa^{\mathrm{mp}}$                & $0.33^{+0.27}_{-0.21}$    & $0.41^{+0.25}_{-0.23}$    & $-0.21^{+0.25}_{-0.25}$   \\
$Z_0^{\mathrm{mr}}$                   & $-1.75^{+0.04}_{-0.04}$   & $-1.71^{+0.05}_{-0.05}$   & $-1.96^{+0.02}_{-0.02}$   \\
$Z_0^{\mathrm{mp}}$                   & $-2.28^{+0.11}_{-0.09}$   & $-2.24^{+0.05}_{-0.06}$   & $-2.32^{+0.04}_{-0.04}$   \\
$\sigma_Z^{\mathrm{mr}}$              & $0.24^{+0.03}_{-0.03}$    & $0.27^{+0.04}_{-0.04}$    & $0.26^{+0.02}_{-0.02}$    \\
$\sigma_Z^{\mathrm{mp}}$              & $0.36^{+0.04}_{-0.04}$    & $0.50^{+0.03}_{-0.04}$    & $0.50^{+0.03}_{-0.02}$    \\
$h_1$                                 & $0.07^{+0.16}_{-0.05}$    & $0.07^{+0.09}_{-0.05}$    & $0.03^{+0.05}_{-0.02}$    \\
$h_2$                                 & $0.91^{+0.07}_{-0.15}$    & $0.95^{+0.04}_{-0.09}$    & $0.89^{+0.08}_{-0.13}$    \\
\hline
\end{tabular}
\caption{Best constrained model parameters for the chemodynamical model of three dSphs. The best-fit values are the medians of MCMC post-burn distributions, and errors are the 16th and 84th percentiles. The meanings of different model parameters can be found in Section~\ref{sec:method}.}
\label{tab:parameters2}
\end{table}

\begin{figure*}
    \centering
    \includegraphics[width=\textwidth]{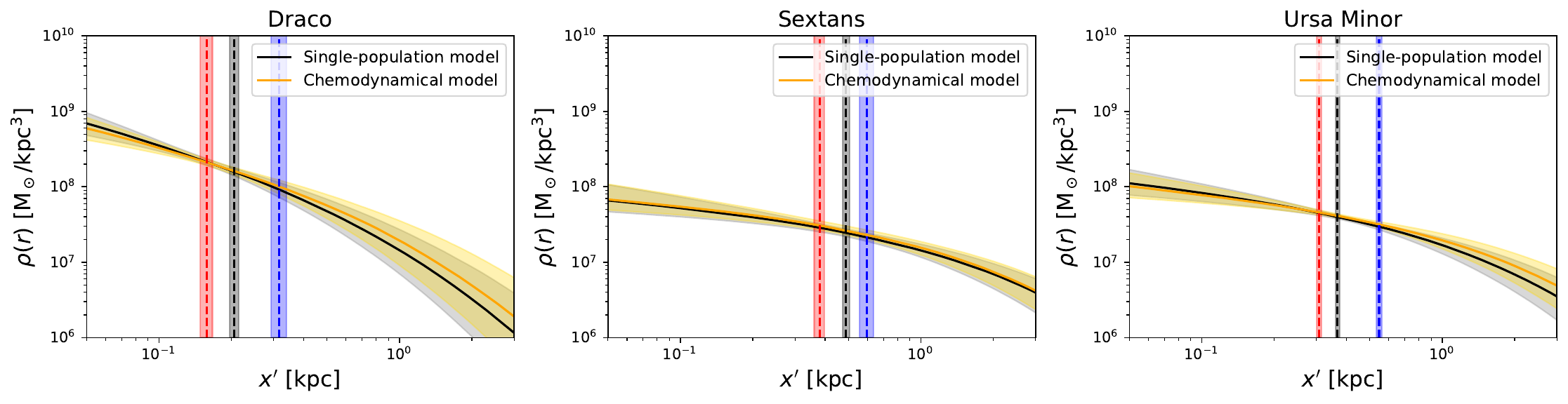}
    \caption{Best recovered dark matter density profiles of three dSphs: (a) Draco, (b) Sextans and (c) UMi. The black and orange solid curves represent median values obtained through our single-population and chemodynamical models, respectively, with shaded regions denoting the 1$\sigma$ confidence regions. The black, red, and blue dashed vertical lines correspond to the half-number radii of the total, metal-rich, and metal-poor populations in each galaxy, with shaded regions denoting the 1$\sigma$ uncertainties. Note that the half-number radii for the metal-rich and metal-poor populations are based on the joint posterior probability of the best model.}
    \label{fig:density profile}
\end{figure*}

We show the best constrained dark matter density profiles through both single-population (black) and chemodynamical (orange) models in Figure \ref{fig:density profile}. Given the consistency in the best constrained model parameters in Figure \ref{fig:para compare}, the best-fit dark matter density profiles by the two models are also fully consistent with each other. Dark matter density profiles are best constrained around the half-number radii of tracer populations (dashed vertical lines), where the associated statistical errors indicated by the shaded regions around density profiles are the smallest. This agrees with the consensus that dynamical modeling generally imposes strongest constraints on the total mass enclosed in the half-number radius of tracers \citep[e.g.][]{Wolf2010,Walker2011,Wang2015,Han2016a,Han2016b,Genina2020,Wang2022,Li2022,Shi2024}.

According to Figure \ref{fig:density profile}, it can be seen that the half-number radii of the metal-rich and metal-poor populations sit on either sides of the half-number radii of the total population, with the metal-rich population having smaller half-number radii. In principle, chemodynamical modeling with two distinct populations is expected to have tighter constraints at two different half-number radii of the two underlying populations, hence leads to better constraints, but Figure \ref{fig:density profile} shows that the best constraints from single-population and chemodynamical models in our analysis are consistent within the errors. This could be limited by the number of available tracer stars used and the relatively large statistical errors of the density profiles, but at least, the agreement between the best constraints of our single-population and two-population chemodynamical models show that our results are self-consistent.

Among the three dSphs, Draco has the steepest inner density profile. It favors an inner dark matter density profile $\gamma=1$ within its $1\sigma$ confidence intervals, which is consistent with cusped NFW profile. On the other hand, dark matter halos of Sextans and UMi have lower central densities and their inner profiles are closer to cores. Across the three dSphs studied in this paper, the inner density slope $\gamma$ of the dark matter halo features a diversity rather than a universal value. We will back discuss this point further in Section \ref{sec:disc} below.

\subsection{Spatial, metallicity and velocity distributions of the metal-rich and metal-poor populations in the chemodynamical model}

Using Equation \ref{eq:relative probability}, for each star in a given dSph, we calculate the probability for it to be in the metal-rich population. Stars that have $P_i^{\prime \, \mathrm{metal-rich}} > 0.5$ and $P_i^{\prime \, \mathrm{metal-poor}} = 1-P_i^{\prime \, \mathrm{metal-rich}} > 0.5$ are classified as members of the metal-rich population and the metal-poor population, respectively.

\begin{figure*}
    \centering
    \begin{minipage}{0.32\textwidth}
        \centering
        \includegraphics[width=\linewidth]{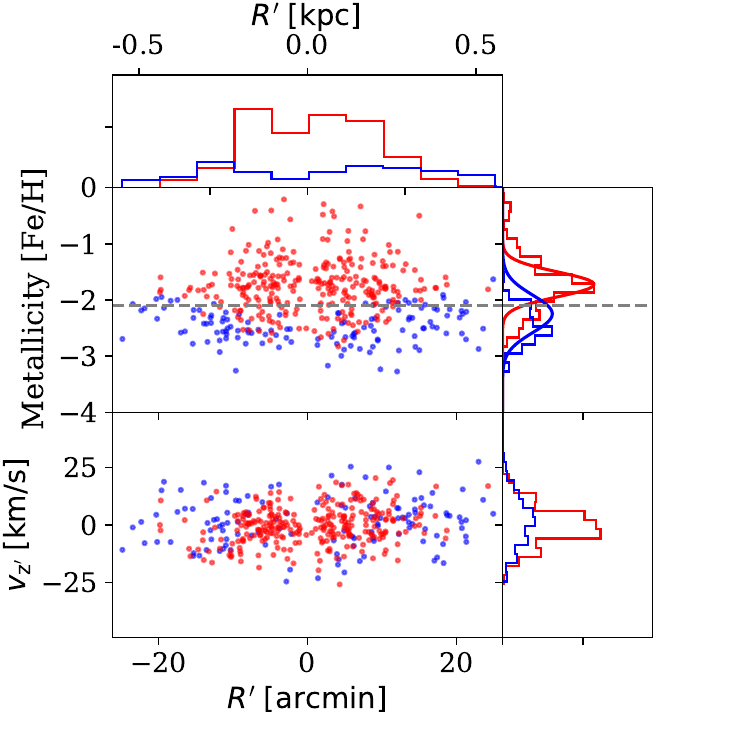}
        \text{(a) Draco}
    \end{minipage}
    \hfill
    \begin{minipage}{0.32\textwidth}
        \centering
        \includegraphics[width=\linewidth]{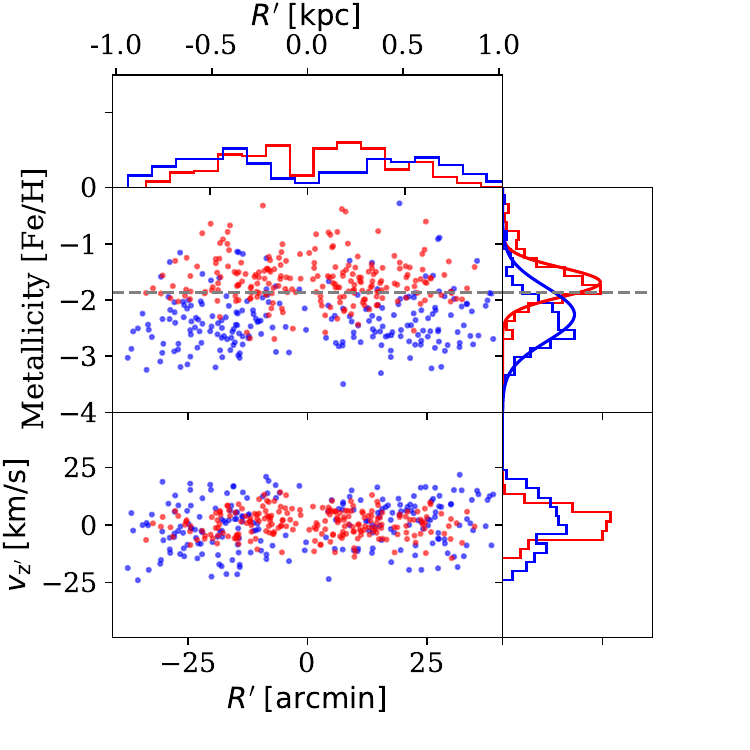}
        \text{(b) Sextans}
    \end{minipage}
    \hfill
    \begin{minipage}{0.32\textwidth}
        \centering
        \includegraphics[width=\linewidth]{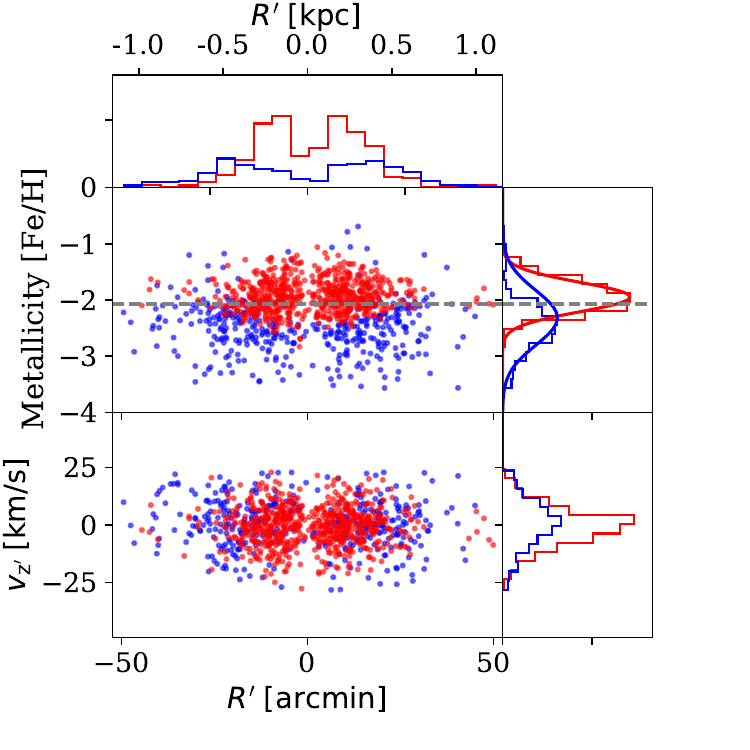}
        \text{(c) UMi}
    \end{minipage}
    \caption{The spatial, metallicity and velocity distributions for the metal-rich (red) and metal-poor (blue) populations recovered by the chemodynamical model. Each sub plot has five panels. {\bf Top histogram:} the spatial distributions of two populations. {\bf Middle-left scatter panel:} $[\mathrm{Fe}/\mathrm{H}]$ versus the projected radius $R^\prime$. {\bf Middle-right histogram:} metallicity distributions of two populations. {\bf Bottom-left scatter panel:} the LOSV ($v_z^\prime$) of each star, versus the projected radius $R^\prime$. {\bf Bottom-right histogram:} the global velocity distributions of two populations. The division of the two populations are based on  Equation~\ref{eq:relative probability}, with metal-rich and metal-poor populations having this probability greater than 0.5. The histogram distributions in the two right panels are for the data, while solid smooth curves with the same colors are model predictions. Grey dashed horizontal lines denote the hard cuts in metallicity to create the red and blue spatial templates.}
    \label{fig:metal}
\end{figure*}

In Figure \ref{fig:metal} we show the spatial, metallicity and velocity distributions of stars in metal-rich and metal-poor populations. The lack of stars at innermost radii is due to the incompleteness of our spectroscopic samples in these regions. Obviously, metal-rich populations are more concentrated in all three dSphs, while metal-poor populations tend to be more extended out to larger radii. These features are consistent with those of template maps in Figure \ref{fig:2sp surface number density}. Similar findings of distinct spatial distributions of two populations in dSphs are also reported in previous works \citep[e.g.][]{Walker2011,Zhu2016,Pace2020}, in that the central part of dSphs may have experienced more recent star formations, and thus the younger metal-rich population is more centrally concentrated.

The metallicity distributions show that the separation of two populations in three dSphs are mostly, but not entirely, determined by hard cuts in metallicity (gray dashed horizontal lines), as the total probability of a star within a population in Equation \ref{eq:probability} is also contributed by the dynamical probability. There are overlaps between the metallicity distributions of the two populations, with the low metallicity tail of the metal-rich population extending lower than the hard cut, and the high metallicity tail of the metal-poor population extending higher than the hard cut. Among the three dSphs, UMi has the most dominant metal-rich population, whereas Sextans features two populations with more comparable sizes. 

\begin{figure*}[htbp!]
    \centering
    \begin{minipage}[b]{0.45\textwidth}
        \includegraphics[width=\linewidth]{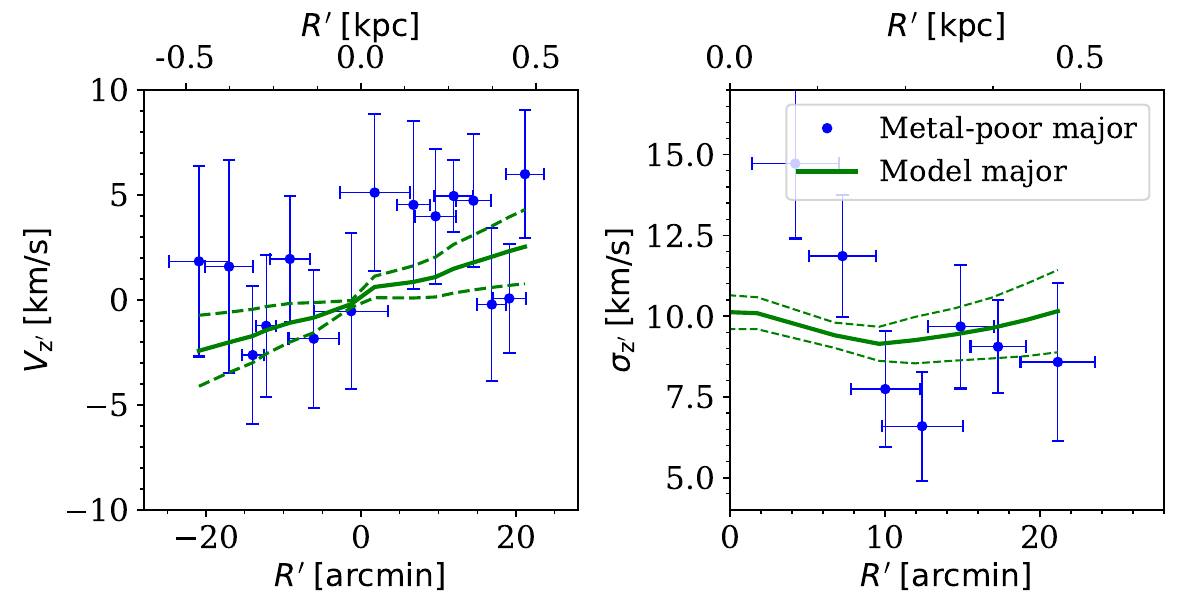}
    \end{minipage}%
    \begin{minipage}[b]{0.45\textwidth}
        \includegraphics[width=\linewidth]{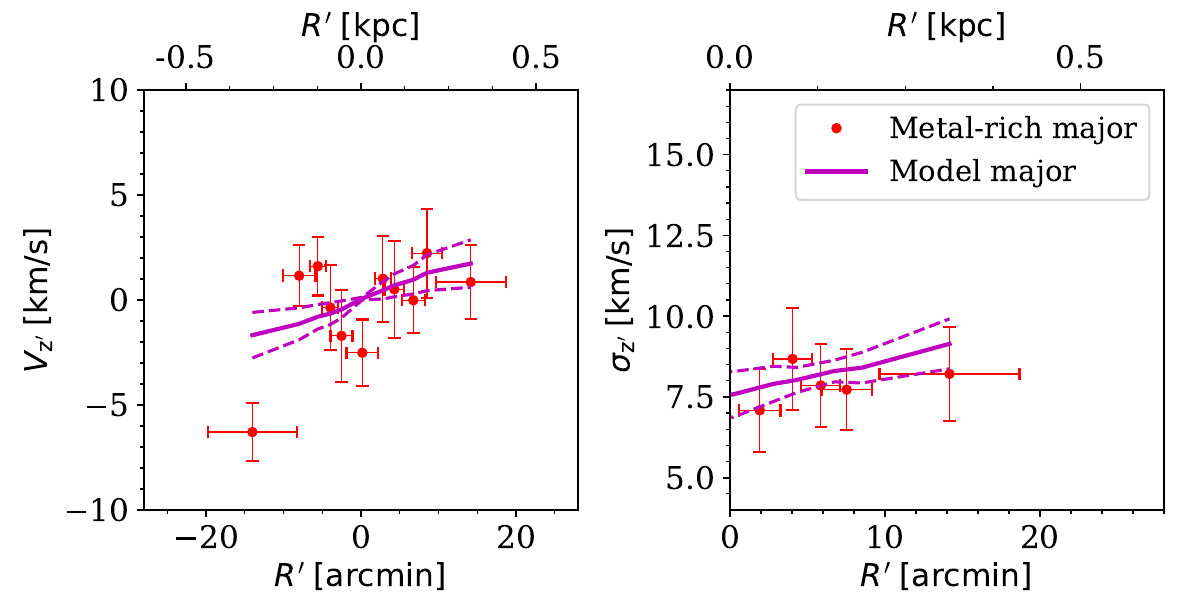}
    \end{minipage}
    \parbox{\textwidth}{\centering \small (a) Draco}

    \begin{minipage}[b]{0.45\textwidth}
        \includegraphics[width=\linewidth]{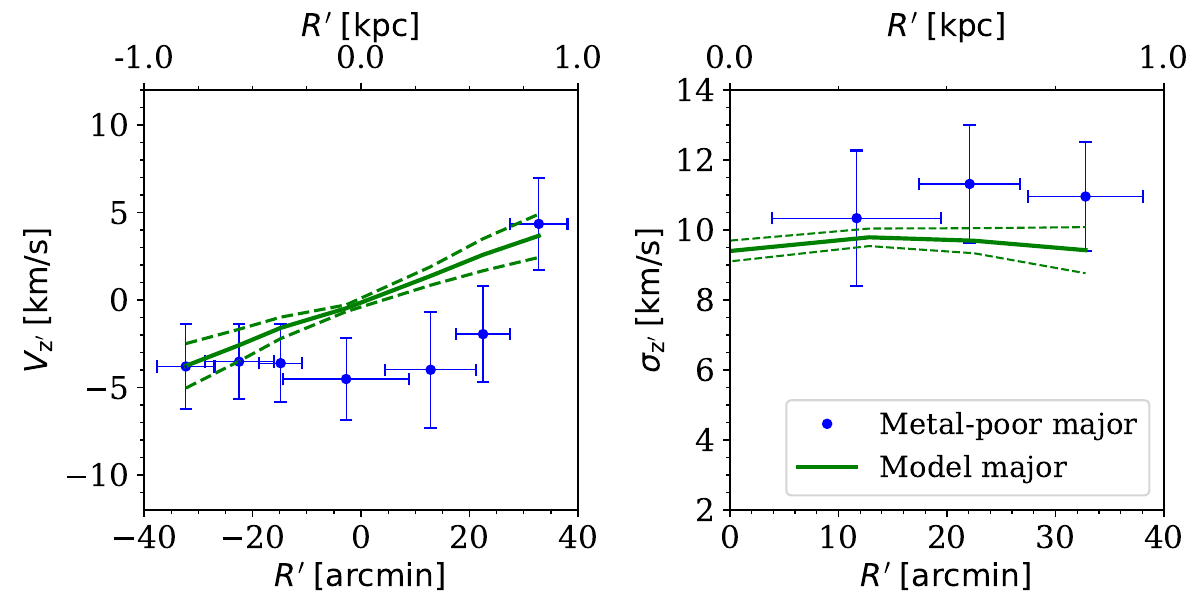}
    \end{minipage}%
    \begin{minipage}[b]{0.45\textwidth}
        \includegraphics[width=\linewidth]{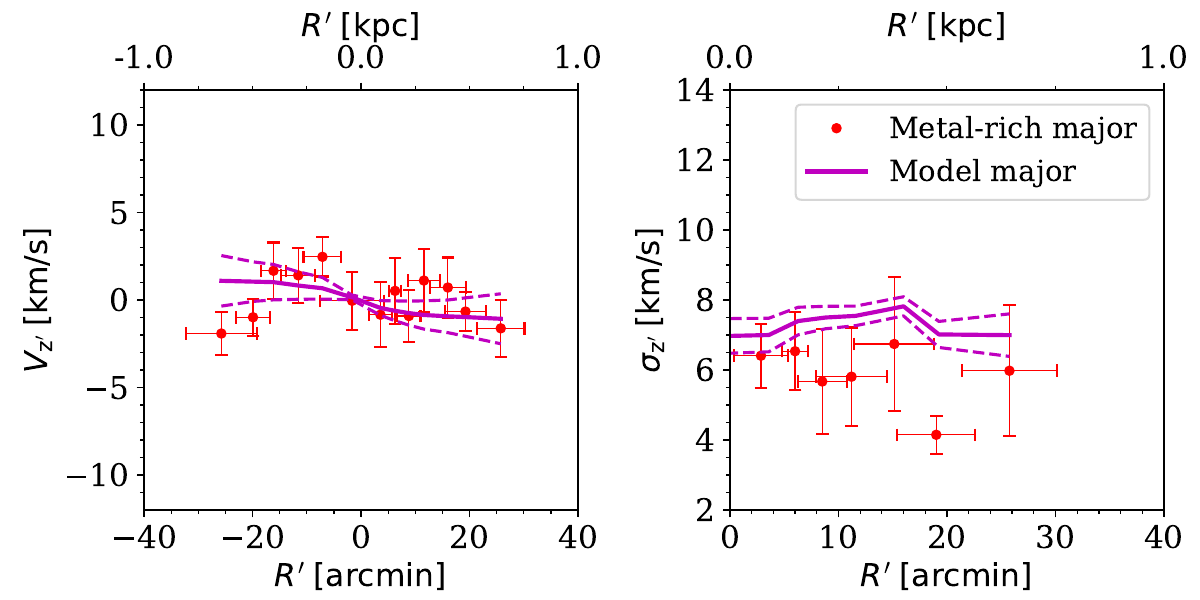}
    \end{minipage}
    \parbox{\textwidth}{\centering \small (b) Sextans}

    \begin{minipage}[b]{0.45\textwidth}
        \includegraphics[width=\linewidth]{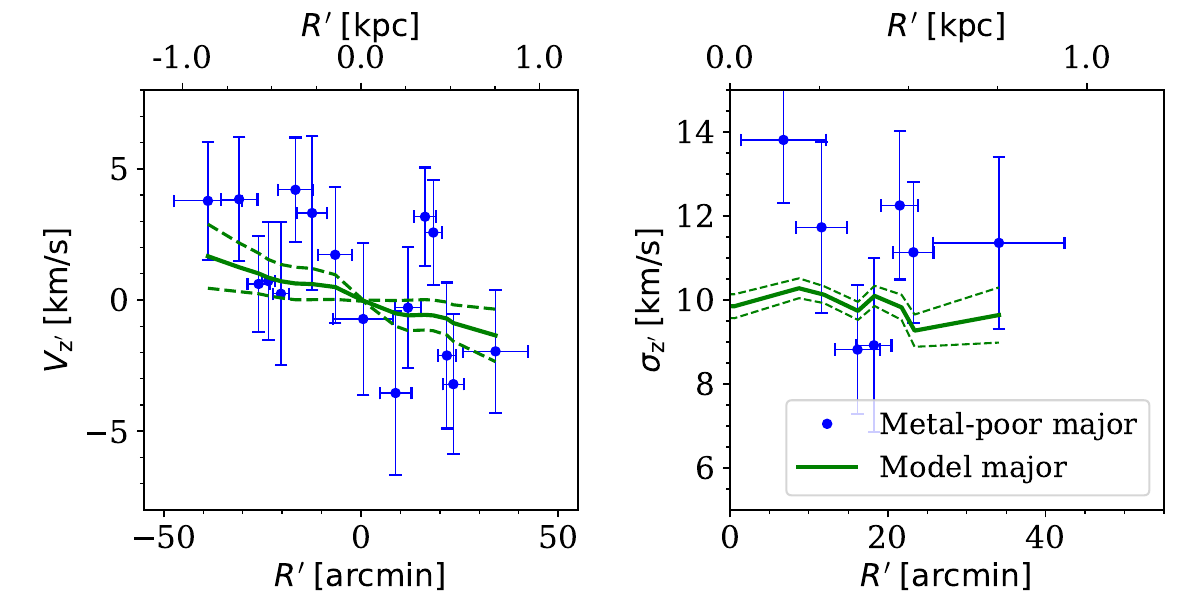}
    \end{minipage}%
    \begin{minipage}[b]{0.45\textwidth}
        \includegraphics[width=\linewidth]{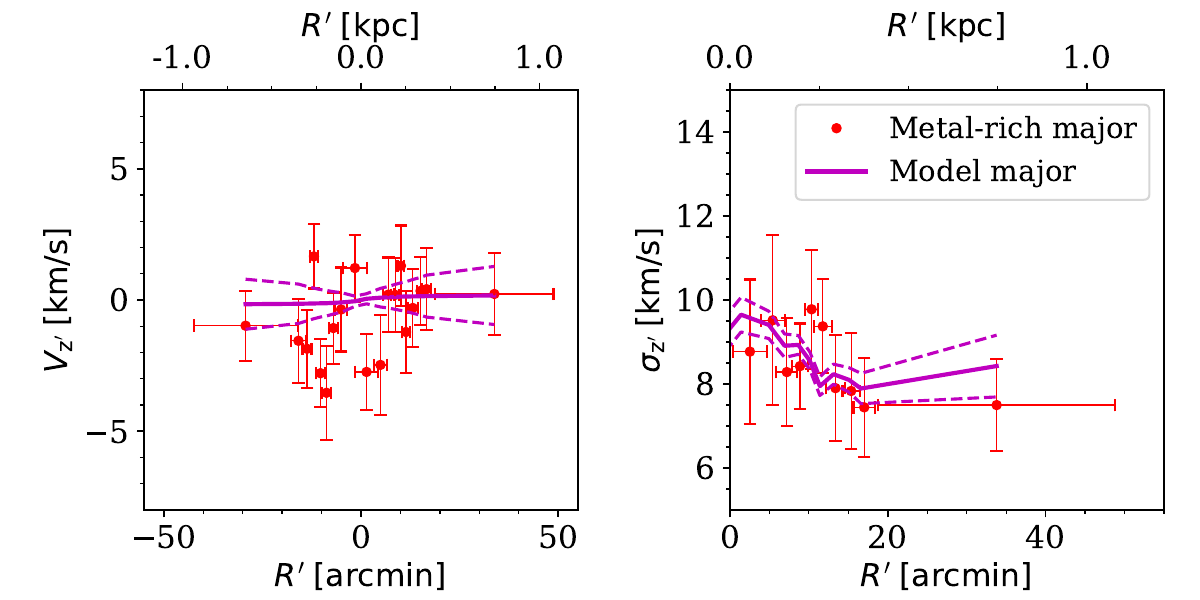}
    \end{minipage}
    \parbox{\textwidth}{\centering \small (c) UMi}

    \caption{As Figure \ref{fig:vz_1sp}, but only for velocity moment profiles along major axes of different stellar populations in three dSphs. Two left and two right columns are the mean LOSV and LOSV dispersion profiles binned along the major axis for the metal-poor (blue dots) and metal-rich (red dots) populations, respectively. We do not show results along the minor axis, as the main trends and comparisons are very similar. In each panel, green and magenta solid curves are the best model predictions (see the legend), with the two dashed curves on both sides showing the 1-$\sigma$ confidence regions. The number of stars in each bin is the same for a given population in one galaxy, but varies between populations and among dSphs.}
    \label{fig:vz_2sp}
\end{figure*}

The two populations also differ in their global velocity distributions. Figure \ref{fig:vz_2sp} more clearly shows the first (mean LOSV) and second (LOSV dispersion) velocity moments of the two populations, similarly to Figure \ref{fig:vz_1sp} for the single-population model, but the two populations are shown in the two left (metal-poor) and two right (metal-rich) columns separately. Here we only show velocity moment profiles along major axes.

For Draco, the two populations identified by the best-fit models (symbols with error bars) show similar mean velocity profiles, which indicate rotations around the minor axes but with different amplitudes. The best-fit model predicts different signs of rotations for the two populations in Sextans, which was also seen in Sculptor \citep{Zhu2016}. For UMi, rotation is only seen in the metal-poor population. However, the two populations show distinct velocity dispersion profiles in all three dSphs. The metal-poor populations are usually kinematically hotter, with negative gradients in their dispersion profiles that decrease with radius. On the other hand, velocity dispersion profiles of the metal-rich populations are kinematically colder and more flattened. The distinct velocity dispersions of the two populations are also consistent with previous studies \citep[e.g.][]{Pace2020}, in that the later formed metal-rich population may have more coherent motions, hence the velocity dispersions are decreased. The older metal-poor population is more relaxed.

In general, the predicted mean velocity profiles by the best constrained model agree well with the data. Unfortunately, the velocity dispersion profiles are poorly fit for the metal-poor populations of both Draco and UMi, and for both populations of Sextans. The same is seen in the metal-poor population of Sculptor in \cite{Zhu2016}. These poor fits are mainly caused by the large difference between the velocity dispersions of the two populations. For all three dSphs, the central velocity dispersions of the metal-poor populations are nearly a factor of two larger than those of the metal-rich ones. If the velocity anisotropy parameters, $\lambda$, for different populations differ, it may cause some differences in the velocity dispersions. But we have tested different $\lambda$ to predict the velocity dispersions using Jeans equations, it is difficult to reproduce a difference as large as a factor of two. The significantly higher velocity dispersion in metal-poor populations cannot be due to the contamination by RR Lyrae stars, as we have excluded the entire horizontal branch (see Section~\ref{sec:data}). Moreover, the effect of binaries cannot account for it either. Among the three dSphs, UMi has multi-epoch observations by DESI for more than six hundred member stars, with a time baseline spanning one year. We have excluded stars with large LOSV variations, defined as the absolute LOSV change between two epoches of observations being greater than a factor 3 of the LOSV error, but the resulting velocity dispersion profiles are very similar to the original ones. We have also checked that the different LOSV dispersions of the two populations cannot be due to their different LOSV errors.

The nearly factor of two difference in the velocity dispersions of the two populations may reflect that the dSph systems deviate from the steady-state assumption behind the Jeans equations, as for multiple tracer populations in the same potential, if all tracers are in steady states, we do not expect such prominent differences in their velocity dispersions. Recent {\it Gaia} observations have also indicated that many MW dSphs may be out of equilibrium \citep{Hammer2023,Hammer2024-1,Hammer2024-2}. The steady-state Jeans solutions cannot reproduce such a prominent difference between different populations in the same potential. Hence it is difficult to simultaneously achieve ideal fits in the observed velocity dispersion profiles of both populations.

For Draco and UMi, the metal-rich populations in the rightmost panels dominate in sample size (see Figure~\ref{fig:metal} above), and thus the models tend to match better the metal-rich population. For Sextans, the metal-rich and metal-poor populations are more comparable in sample size, and the best-fit model tends to reconcile the large difference in velocity dispersion profiles of two populations, resulting in over-estimated dispersions for the metal-rich population, and under-estimated dispersions for the metal-poor population.

Deviations from steady-state assumption can introduce biased constraints. Previously, \citep{Wang2022} applied \textsc{jam} to simulated dwarf galaxies to test the biases. It was found that global expansion/infalling motions can cause deviations from steady states, resulting in over/under-estimated inner densities, but still maintain a good constraint on the total mass within the half-number radii of tracers. In our analysis, we introduce the two-population chemodynamical model because the two distinct half-number radii of the metal-rich and metal-poor populations may help to reduce systematics. Our results are at least self-consistent given the fact that the single-population and chemodynamical models are consistent.

\subsection{Astrophysical factors}

In this subsection, we provide our measurements of the astrophysical factors of Draco, Sextans and UMi, for the purposes of dark matter indirect detections, such as Fermi-LAT \citep{Fermi-LAT}, DAMPE \citep{DAMPE}, HAWC \citep{HAWC}, CTA \citep{CTA} and LHAASO \citep{LHAASO}. The astrophysical factor is determined by the dark matter distribution within the galaxy \citep[e.g.][]{Baltz1999,Bergstrom2006,Geringer-Sameth2015,Hayashi2016,Strigari2018R}. For dark matter annihilation or decay, the astrophysical factor is called the J or D factor, respectively. The J and D factors can be written as

\begin{align}
    J&=\iint \rho_{\mathrm{DM}}^2(l,\Omega) \, \mathrm{d}l\mathrm{d}\Omega, \label{eq:j factor} \\
    D&=\iint \rho_{\mathrm{DM}}(l,\Omega) \, \mathrm{d}l\mathrm{d}\Omega,
    \label{eq:d factor}
\end{align}
where $\rho_{\mathrm{DM}}$ is the dark matter density, $l$ is the distance along the line-of-sight direction, and $\Omega$ is the solid angle centered on the dSph.

\begin{figure}[htbp!]
    \centering
    \begin{minipage}{\linewidth}
        \centering
        \includegraphics[width=0.8\linewidth]{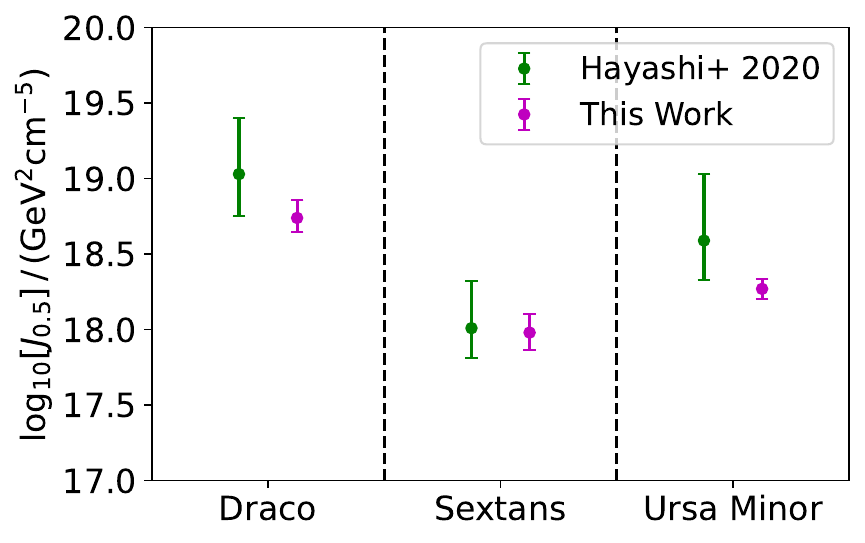}
    \end{minipage}
    \vfill
    \begin{minipage}{\linewidth}
        \centering
        \includegraphics[width=0.8\linewidth]{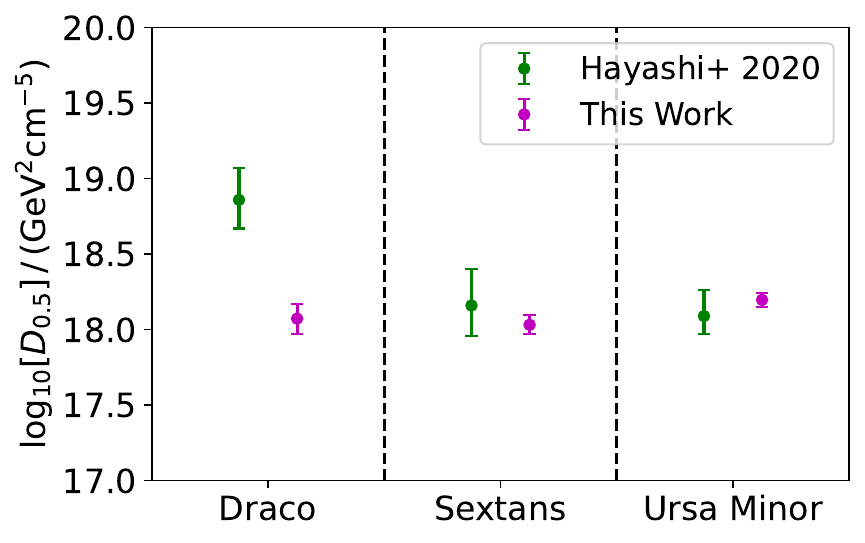}
    \end{minipage}
    \caption{Measured J (top) and D (bottom) factors for the three dSphs. Green and magenta symbols denote those from \cite{Hayashi2020} and this work.}
    \label{fig:jd factor}
\end{figure}

We calculate the J and D factors of the three dSphs with integrals covering a fixed solid angle of $\Delta \Omega=0.5^\circ$, and compare our results with those from \cite{Hayashi2020} in Figure \ref{fig:jd factor}. The differences are mainly due to different dark matter halo models adopted and different truncation radii. \cite{Hayashi2020} assumed axisymmetric dark matter halos with generalized Hernquist profiles \citep{Hernquist1990,Zhao1996}, whereas we consider spherical dark matter halos with generalized NFW profiles, and the outer density slopes are fixed to 4. 
Both \cite{Hayashi2020} and us truncate the dark matter halo at the projected radius of the outermost observed member star for each dSphs. However, stars beyond $2.5$ half-light radius from the center of each dSphs are not used in our analysis, and thus the truncation radius  in our analysis is smaller than that of \cite{Hayashi2020}. In Figure \ref{fig:jd factor}, the most prominent difference between \cite{Hayashi2020} and this work is the D factor for Draco. This is because their truncation radius for Draco is much larger than ours and the dark matter distribution at larger radii contributes more to the D factor than the J factor. Besides, since we truncate at smaller radii where dark matter density profiles are better constrained, uncertainties of our J and D factors are smaller.



\section{Discussion}

\label{sec:disc}

\begin{figure}[h]
    \centering
    \includegraphics[width=0.9\linewidth]{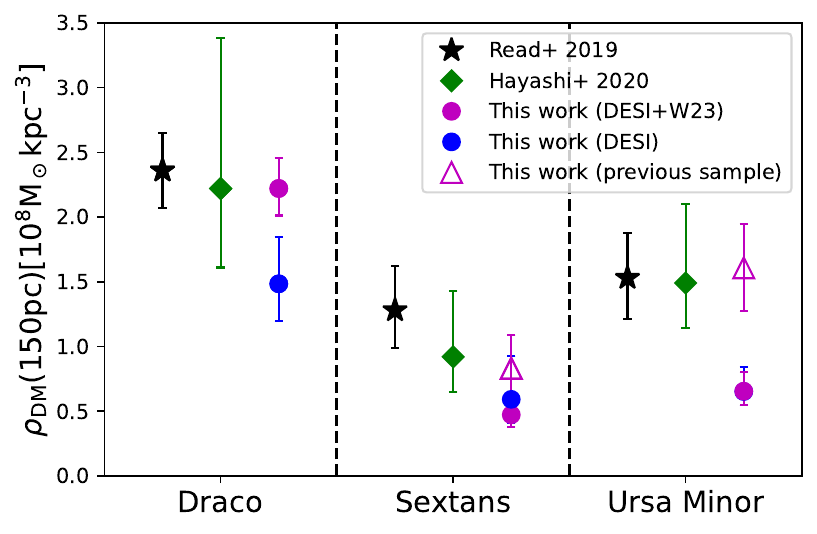}
    \caption{The dark matter densities at $150\mathrm{pc}$ for the three dSphs. The black stars and green diamonds represent results in \cite{Read2019} and \cite{Hayashi2020}. The magenta filled circles represent our results using the combined member star sample from DESI observation and \cite{Walker2023}, based on the best constrained chemodynamical model. Results of the single-population model are very similar (see Figure \ref{fig:density profile}), and thus are not shown. Blue filled circles are measurements based on DESI observed member stars only. For UMi the magenta filled circle overlaps the blue filled circle. We also repeat our analysis for Sextans and UMi using the same member star samples as \cite{Hayashi2020}, and show the results by hollow magenta triangles.}
    \label{fig:rho at 150pc 3}
\end{figure}

In \cite{Read2019}, the dark matter density at 150~pc, $\rho_{\mathrm{DM}}(150\mathrm{pc})$, is used as a proxy of the inner density slope $\gamma$. We compare our results of $\rho_{\mathrm{DM}}(150\mathrm{pc})$ with those reported by \cite{Read2019} and \cite{Hayashi2020}\footnote{We use the updated values kindly provided by Kohei Hayashi, which are different from the values listed in Table 2 of his work.} in Figure \ref{fig:rho at 150pc 3}. For the three dSphs used in our analysis, we mark our best recovered $\rho_{\mathrm{DM}}(150\mathrm{pc})$ from the chemodynamical model above by the magenta filled circles, and we do not repeatedly show those for the single-population model as they are very similar (see Figure \ref{fig:density profile}). Previous measurements by \cite{Read2019} and \cite{Hayashi2020} are plotted with black stars and green diamonds.

For Draco, our measured $\rho_{\mathrm{DM}}(150\mathrm{pc})$ agrees well with both \cite{Read2019} and \cite{Hayashi2020}. For Sextans, three measurements are different from each other, though the measurements of \cite{Hayashi2020} and \cite{Read2019} still agree within the errors. This discrepancy is very likely due to the fact that both \cite{Hayashi2020} and us adopt axisymmetric Jeans analysis, while \cite{Read2019} used spherical Jeans models. Thus our measurement and the measurement by \cite{Hayashi2020} for Sextans are both lower than that of \cite{Read2019}. Besides, the sample of member stars used in \cite{Hayashi2020} is different from that in this work. We have carefully repeated our calculation by using the same sample as in \cite{Hayashi2020}, and we can reproduce the result of \cite{Hayashi2020} for Sextans, as shown by the hollow magenta triangle in Figure \ref{fig:rho at 150pc 3}. For UMi, our result is inconsistent with both \cite{Read2019} and \cite{Hayashi2020}, and the difference is larger than that for Sextans. We have explicitly checked that this disagreement is mainly due to the large difference in the sample of member stars used for UMi, as the member star sample with LOSV measurements in our analysis is three times larger than those in \cite{Hayashi2020} or \cite{Read2019}. We have also repeated our analysis by using the same sample of UMi as in \cite{Hayashi2020}, and we again obtain consistent result with both \cite{Read2019} and \cite{Hayashi2020}, with the best constraint shown by the hollow magenta triangle in Figure \ref{fig:rho at 150pc 3}. This test robustly confirms that it is mainly the difference in the sample of member stars used in different analyses that leads to the discrepancy between our measurement and those in the two previous studies for UMi.

We have checked that the intrinsic velocity dispersion profiles for the member star sample of \cite{Hayashi2020} and of our analysis above (DESI $+$ \cite{Walker2023}) differ for both Sextans and UMi, and in turn significantly influence the inferred dark matter content. Here the intrinsic velocity dispersions deconvolved from observational errors are calculated following the method\footnote{They use a two-parameter Gaussian likelihood function similar to that of \cite{Walker2006} to calculate the systemic velocity and the intrinsic velocity dispersion of a sample of stars: $\log \mathcal{L} = -\frac{1}{2} \left[ \sum \log(\sigma_{v_{\text{sys}}}^2 + \sigma_{v_i}^2) + \sum (v_i - v_{\text{sys}})^2 / (\sigma_{v_i}^2 + \sigma_{v_{\text{sys}}}^2) \right]$, where $v_{\text{sys}}$ and $\sigma_{v_{\text{sys}}}$ are the systemic velocity and the intrinsic velocity dispersion of the sample, and $v_i$ and $\sigma_{v_i}$ are the velocity and the velocity uncertainty of each star.} in \cite{Li2017}, and we have corrected LOSV offsets between the two data samples. We find that the sample in \cite{Hayashi2020} has higher intrinsic velocity dispersions in the central region than our sample, and this is the reason for the lower density $\rho_{\mathrm{DM}}(150\mathrm{pc})$ obtained in our analysis. In fact, the member stars of UMi in \cite{Hayashi2020} are more metal-poor than our sample, which results in higher velocity dispersion in the center. However, given the significantly increased number of DESI observed member stars for UMi, we believe our sample is more complete.

The different intrinsic velocity dispersions are due to the difference in the selection functions of these observations. Throughout this paper, we have been assuming that the fiber assignment does not depend on velocity, and the velocity dispersion profiles of observed member stars in the three dSphs do not differ from those of targets. However, based on our discussion above, this assumption does not hold for all observations, i.e., the sample of stars used in \cite{Hayashi2020} is more metal-poor than ours. 

Since different samples may subject to different selection functions that can affect the velocity dispersion profiles, the combination of two data sets in this work, i.e., data from DESI and \cite{Walker2023}, may also introduce complicated selection effects. Thus we perform further checks. For each dSph, we first calculate the intrinsic velocity dispersions for the same sample of stars existing in both DESI and \cite{Walker2023}. The intrinsic velocity dispersions computed using LOSV measurements from both fully agree with each other, indicating the measurements of LOSV and associated uncertainties are reliable in both DESI and \cite{Walker2023}. However, if we use all member stars either in DESI or \cite{Walker2023}, after correcting for the mean offset between LOSVs in the two data sets, the resulting intrinsic velocity dispersion profiles based on the data set of \cite{Walker2023} are slightly different from those of DESI observed member stars for Draco and Sextans, while there is no difference for UMi. The agreement between DESI and \cite{Walker2023} UMi sample is reassuring, indicating that the large sample of DESI observed UMi member stars is unlikely subject to strong selection biases.

Given the difference in the intrinsic velocity dispersions based on DESI and \cite{Walker2023} member stars for Draco and Sextans, we repeat our modeling using only DESI data for them. 
For Draco and Sextans, the best constrained dark matter density profiles slightly differ when we use only DESI observed member stars, and when using the combined sample of DESI $+$ \cite{Walker2023}. Nevertheless, the differences are still marginally consistent within the large 1-$\sigma$ errors.

The best recovered $\rho_\mathrm{DM}(150\mathrm{pc})$ of three dSphs using only DESI data are shown by the blue filled circles in Figure~\ref{fig:rho at 150pc 3}. Using only DESI observed stars yields higher intrinsic velocity dispersions in the central region for Sextans whereas lower ones for Draco than using the combined data, hence resulting in higher and lower $\rho_\mathrm{DM}(150\mathrm{pc})$, respectively. For Sextans, since the difference between the blue and magenta circles  is less significant compared with the errorbars, we think selection effects associated with Sextans are not strong enough to violate the conclusion. For Draco, DESI is missing a lot of metal-poor stars in the central region, which is at least partially because the number of observed member stars in DESI is small (less than 100, see Table~\ref{tab:dwarf prop}). On the other hand, \cite{Walker2023} observes more member stars in the center of Draco. Hence the best constrained $\rho_\mathrm{DM}(150\mathrm{pc})$ for Draco is more robust with the combined data set of DESI $+$ \cite{Walker2023}, i.e., the magenta circle in Figure \ref{fig:rho at 150pc 3}.

\begin{figure*}[htbp!]
    \centering
    \includegraphics[width=0.65\textwidth]{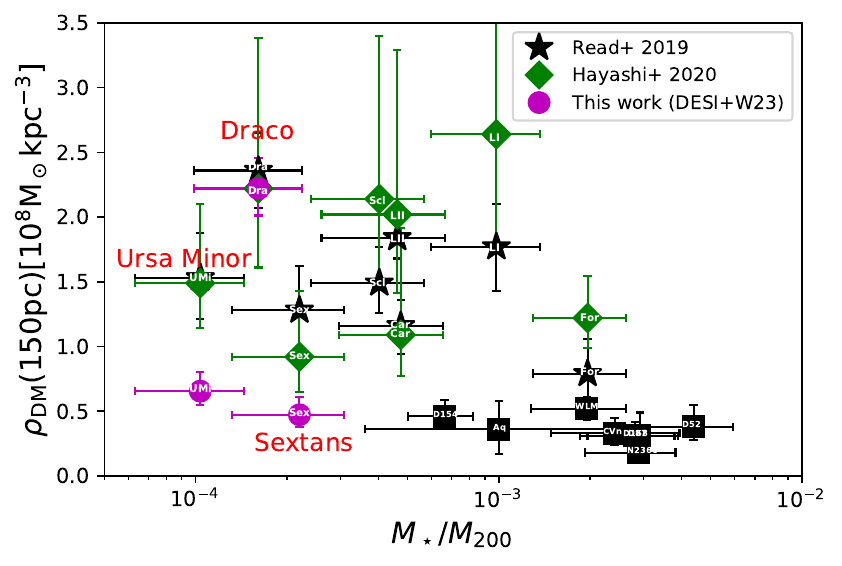}
    \caption{The dark matter densities at $150\mathrm{pc}$ ($y$-axis) versus the stellar-to-halo mass ratios ($x$-axis) for dwarf galaxies in a few previous studies. The black filled squares and stars denote results in \cite{Read2019} for dwarf irregular galaxies (dIrrs) and dwarf spheroidal galaxies (dSphs), respectively. The green symbols represent results in \cite{Hayashi2020}. The magenta filled circles represent results in our main analysis using the combined member star sample from DESI observation and \cite{Walker2023}, based on the best constrained chemodynamical model in this work. No obvious correlation is observed.}
    \label{fig:rho at 150pc all}
\end{figure*}

Finally, we show in Figure~\ref{fig:rho at 150pc all} the best recovered $\rho_\mathrm{DM}(150\mathrm{pc})$ for dwarf galaxies in \cite{Read2019} and \cite{Hayashi2020}, in addition to the three dSphs in our analysis. A few dwarf irregular galaxies used by \cite{Read2019} are marked with black squares. The values of $M_\star / M_{200}$ (the $x$-axis) for all dSphs are just taken from \cite{Read2019}, which are obtained through abundance matching. We do not calculate them again. \cite{Read2019} reached a very important conclusion that there is a negative correlation between $\rho_{\mathrm{DM}}(150\mathrm{pc})$ and $M_\star / M_{200}$, which supports stellar feedback as the scenario to explain the core-cusp problem for dwarf galaxies. However, similar trend is not clearly seen in \cite{Hayashi2020}, due to the fact that \cite{Hayashi2020} did not include those dwarf Irregulars used in \cite{Read2019}, and the measured inner densities for Fornax and Leo~I on the right side of Figure \ref{fig:rho at 150pc all} are higher than those by \cite{Read2019}. Our results, denoted by filled magenta circles in Figure \ref{fig:rho at 150pc all}, do not fully follow this anti-correlation due to the lower values of $\rho_{\mathrm{DM}}(150\mathrm{pc})$ of Sextans and UMi.

Even with only three dSphs, we find a diversity of $\rho_{\mathrm{DM}}(150\mathrm{pc})$, or the inner density slope, $\gamma$, with $\rho_{\mathrm{DM}}(150\mathrm{pc})$ spanning nearly a factor of five among the three dSphs. Draco has the highest density $\rho_{\mathrm{DM}}(150\mathrm{pc})$ with the inner density slope $\gamma=0.71^{+0.34}_{-0.35}$ close to $1$, while dark halos of Sextans and UMi are less dense in inner regions with smaller $\gamma$, $0.26^{+0.22}_{-0.12}$ and $0.33^{+0.20}_{-0.16}$, which are closer to cored models. Moreover, by comparing the measured inner densities with \cite{Read2019} and \cite{Hayashi2020}, and by repeating our analysis using member stars observed in different surveys, we have demonstrated that the currently best constrained inner density slopes are still sensitive to the model assumptions and to sample selection effects.

\section{Conclusions}
\label{sec:concl}

Three classical dSphs (Draco, Sextans and UMi) have been observed by DESI MWS so far. We combine the line-of-sight velocity and metallicity measurements from DESI MWS with those in \cite{Walker2023} to construct the so-far largest member star samples with spectroscopic observations of the three dSphs. To recover their inner dark matter distributions, the axisymmetric Jeans Anisotropic Multi-Gaussian Expansion modeling (\textsc{jam}) approach is adopted. In particular, we use both traditional single-population Jeans modeling and multiple population chemodynamical modeling for our analysis.

We find the best recovered model parameters and the underlying dark matter radial density profiles are consistent between single-population and chemodynamical models. The dark matter density profiles are best constrained around the half-number radii of tracer member stars. A diversity is found for the inner density slopes $\gamma$ of host dark matter halos. The inner density slopes are $0.71^{+0.34}_{-0.35}$, $0.26^{+0.22}_{-0.12}$ and $0.33^{+0.20}_{-0.16}$ for Draco, Sextans and UMi, respectively.

With the chemodynamical model, member stars with spectroscopic observations of each dSph are divided into metal-rich and metal-poor populations, based on their chemical, spatial and dynamical properties. The metal-rich populations are more centrally concentrated, while the metal-poor populations are more extended. Moreover, the metal-rich populations are dynamically colder, featuring lower velocity dispersion profiles than the metal-poor populations in all three dSphs. We identify large differences between the velocity dispersions of the metal-rich and metal-poor populations, likely indicating deviations from the steady-state model assumption.

We identify rotations in all three dSphs. In particular, the metal-rich and metal-poor populations in Draco show the same sign of rotation around the minor axes but with different amplitudes. The two populations in Sextans show different signs of rotations. For UMi, rotation is only seen in the metal-poor population.

We still observe some correlations between the velocity anisotropy parameter and the potential parameters, indicating the existence of mass-anisotropy degeneracy in our results. The {\it Gaia} proper motions are subject to very large uncertainties at the distances of our three dSphs, and thus including or not including {\it Gaia} proper motions lead to almost the same constraints. The mass-anisotropy degeneracy can be broken with more precise proper motion measurements in the future, or with higher order Jeans analysis \citep[e.g.][]{Merrifield1990,Lokas2002,Richardson2013,Wardana2025}.

We compare our results with those in two important previous studies, \cite{Read2019} and \cite{Hayashi2020}. Our measurement for Draco agrees well with both of \cite{Read2019} and \cite{Hayashi2020}. For Sextans, all three measurements are different from each other. This discrepancy is mainly because both \cite{Hayashi2020} and this study adopt axisymmetric Jeans analysis, whereas the model of \cite{Read2019} is based on spherical Jeans modeling. Besides, the sample of member stars used by \cite{Hayashi2020} has higher velocity dispersion in the center, resulting in higher dark matter density. Our best recovered $\rho_\mathrm{DM}(150)$ of UMi is much lower than both \cite{Read2019} and \cite{Hayashi2020}, which is because the UMi member stars in \cite{Read2019} and \cite{Hayashi2020} are more metal-poor and have higher velocity dispersion in the center than ours. Since the member star sample observed by DESI is significantly larger than those in previous observations, especially for UMi, we believe our sample subjects less to selection effects.

In addition to UMi, we also see possible sample selection effects for Draco and Sextans, i.e., their intrinsic velocity dispersions based on the sample of member stars in \cite{Walker2023} are different from those based on only DESI observed member stars. We thus compare the best constraints of using only DESI member stars with the DESI $+$ \cite{Walker2023} data. For Sextans, the best constrained $\rho_\mathrm{DM}(150\mathrm{pc})$ by using DESI data is slightly higher than that of using the combined data set, but the difference is less significant compared with errors. For Draco, if only using DESI observed member stars, we find $\rho_\mathrm{DM}(150\mathrm{pc})$ becomes significantly lower than that of using DESI $+$ \cite{Walker2023} data, mainly due to the the lacking of DESI observed metal-poor member stars in central regions. 

In particular, \cite{Read2019} finds an anti-correlation between the inner dark matter densities and the stellar-to-halo mass ratios for dSphs, which supports the stellar feedback scenario to explain the core-cusp problem. The measurements by \cite{Hayashi2020} and in our current analysis do not fully track the anti-correlation reported by \cite{Read2019} at the low value end of $M_\ast/M_{200}$. Our results indicate that the study of the dark matter content of dSphs through stellar kinematics is still subject to uncertainties in both the model assumption and selection effects in the observed data.

We also estimate astrophysical J and D factors of the three dSphs. The data used for all figures in this work, including the best-fit inner densities and the J and D factors, can be found at \url{https://zenodo.org/records/16919342}.

\section*{Acknowledgments}

This work is supported by NSFC (12273021, 12022307, 12573022) and the National Key R\&D Program of China (2023YFA1605600, 2023YFA1605601, 2023YFA1607800, 2023YFA1607801), 111 project (No.\ B20019), and the science research grants from the China Manned Space Project (No.\ CMS-CSST-2021-A03). We thank the sponsorship from Yangyang Development Fund. The computations of this work are carried on the Gravity supercomputer at the Department of Astronomy, Shanghai Jiao Tong University and the National Energy Research Scientific Computing Center (NERSC). HY is supported by T.D. Lee scholarship. LZ acknowledges the support from CAS Project for Young Scientists in Basic Research, Grant No. YSBR-062. SK acknowledges support from the Science \& Technology Facilities Council (STFC) grant ST/Y001001/1.

We are grateful for helps by Kohei Hayashi, who has kindly shared with us the member star catalog and the updated results of his work. We also thank our DESI publication handler, Ting Tan, for his helps and coordinations.

This material is based upon work supported by the U.S. Department of Energy (DOE), Office of Science, Office of High-Energy Physics, under Contract No. DE–AC02–05CH11231, and by the National Energy Research Scientific Computing Center, a DOE Office of Science User Facility under the same contract. Additional support for DESI was provided by the U.S. National Science Foundation (NSF), Division of Astronomical Sciences under Contract No. AST-0950945 to the NSF’s National Optical-Infrared Astronomy Research Laboratory; the Science and Technology Facilities Council of the United Kingdom; the Gordon and Betty Moore Foundation; the Heising-Simons Foundation; the French Alternative Energies and Atomic Energy Commission (CEA); the National Council of Humanities, Science and Technology of Mexico (CONAHCYT); the Ministry of Science, Innovation and Universities of Spain (MICIU/AEI/10.13039/501100011033), and by the DESI Member Institutions: \url{https://www.desi.lbl.gov/collaborating-institutions}. Any opinions, findings, and conclusions or recommendations expressed in this material are those of the author(s) and do not necessarily reflect the views of the U. S. National Science Foundation, the U. S. Department of Energy, or any of the listed funding agencies.

The authors are honored to be permitted to conduct scientific research on I'oligam Du'ag (Kitt Peak), a mountain with particular significance to the Tohono O’odham Nation.

For the purpose of open access, the author has applied a Creative
Commons Attribution (CC BY) licence to any Author Accepted
Manuscript version arising from this submission.

\bibliography{master}

\end{document}